# A probabilistic method for the estimation of earthquake source parameters from spectral inversion: application to the 2016-2017 Central Italy seismic sequence

**Authors :** *Mariano Supino[1], Gaetano Festa[1] and Aldo Zollo[1]*

(1) Department of Physics " Ettore Pancini ", Università degli Studi di Napoli " Federico II ", Italy

**Corresponding author :** Mariano Supino; E-mail : mariano.supino@unina.it; Tel. : +39 081679927; Fax : +39 0810093584

**Summary**

We develop a probabilistic framework based on the conjunction of states of information between data and model, to jointly retrieve earthquake source parameters and anelastic attenuation factor from inversion of displacement amplitude spectra. The evaluation of the joint probability density functions (PDFs) enables us to take into account between-parameter correlations in the final estimates of the parameters and related uncertainties. Following this approach, we first search for the maximum of the a-posteriori PDF through the basin hopping technique that couples a global exploration built on a Markov chain with a local deterministic maximization. Then we compute statistical indicators (mean, variance and correlation coefficients) on source parameters and anelastic attenuation through integration of the PDF in the vicinity of the maximum likelihood solution. Definition of quality criteria based on the signal to noise ratio and similarity of the marginal PDFs with a Gaussian function enable us to define the frequency domain for the inversion and to get rid of unconstrained solutions. We perform synthetic tests to assess theoretical correlations as a function of the signal to noise ratio and to define the minimum bandwidth around the corner frequency for consistent parameter resolution.

As an application, we finally estimate the source parameters for the 2016-2017 Central Italy seismic sequence. We found that the classical scaling between the seismic moment and the corner frequency holds, with an average stress drop of $\Delta\sigma = 2.1 \pm 0.3\, MPa$. However, the main events in the sequence exhibit a stress drop larger than the average value. Finally, the small seismic efficiency indicates a stress overshoot, possibly due to dynamic effects or large frictional efficiency.

# 1. Introduction

The characterization of the source parameters of small to moderate earthquakes is an important step in understanding the general mechanisms of earthquake nucleation and propagation, since it shines a light on the physical processes involving faults over different space and time scales. The major issue in understanding the physics of such ruptures is the correct characterization of the energy budget associated with the different mechanisms which take place during the earthquake nucleation, unstable propagation, high frequency radiation and arrest. Although the seismic rupture non-linearly combines several space and time scales, a few macroscopic parameters can provide insights on its evolution, such as the earthquake size and the stress drop released during a seismic event. However, their estimation is very uncertain (Cotton *et al.* 2013), owing to uncertainties in data and models and to the coupling between source and wave propagation up to the observation sites. Indeed, actual estimates of stress drop and apparent stress do not allow to distinguish if these parameters are independent of or they scale with the earthquake size (e.g. Scholz 1994; Ide & Beroza 2001; Shaw 2009; Cocco *et al.* 2016).

Different kinematic and dynamic source models, built on circular cracks, have been proposed to infer source parameters of small to moderate earthquakes from observations. They are derived from different initial hypotheses, such as an instantaneously stress pulse applied on the fault (Brune 1970), a rupture evolving at constant velocity (Sato & Hirasawa 1973) or a rupture spontaneously propagating under pre-defined stress conditions (Madariaga 1976), but they all provide a description of the far-field displacement spectrum coherent with observations. In these models, the displacement amplitude spectrum is almost constant at low-frequencies and proportional to the seismic moment, and decays as $f^{-2}$ at large frequencies. The cut-off frequency of this spectrum is related to the source radius and can be connected to the average stress drop using the static solution of Eshelby (1957) and Keilis-Borok (1959).

Several methods for the characterization of earthquake sources using a spectral analysis have been proposed in literature. Parametric approaches provide a representation of the source and propagation through a limited set of parameters. Single station solutions make use of theoretical Green's functions with a frequency-independent attenuation factor (Boatwright *et al.* 1991; Abercrombie 1995). To reduce the epistemic uncertainty related to the wave propagation and site effects, theoretical Green's function can be replaced by records of small earthquakes, playing the role of empirical Green's functions (EGFs); in this case the spectral ratio between couples of stations is modelled (Prieto *et al.* 2004; Abercrombie & Rice 2005). However, EGFs are required to have more than one point of magnitude smaller than the main event to characterize, to resolve source parameters (Abercrombie & Rice 2005). On the contrary, non-parametric techniques do not assume the functional shape of the Green's function, but they remove it from the spectra using a data-driven approach (Castro *et al.* 1990; Oth *et al.* 2007).

Since the stress drop is connected to the cube of the corner frequency, characterization of earthquake scaling laws requires a correct assessment of the uncertainties and correlations among parameters. Confidence intervals for source parameters estimated from single station spectra are usually derived from the fitting procedure, using, for instance, a jackknife analysis (Prieto *et al.* 2007) or by a linearization of the problem (Zollo *et al.* 2014). When using an EGF-based approach, uncertainties also rely on the quality and number of EGFs, and on their location and size with respect to the main event. Uncertainties can be larger than 50% for single EGFs located outside the source radius of the main event (Kane *et al.* 2011; Abercrombie 2015).

Large uncertainties in earthquake source parameters and Q arise from correlations (Abercrombie 1995), and solutions have been proposed with the goal of reducing them, such as a multi-step, iterative approach (Zollo *et al.* 2014). However, a systematic comparison between different methodologies highlighted the dependence of the results on the fitting model (Oye *et al.* 2005; Sonley & Abercrombie 2006).

A probabilistic approach to this inverse problem can allow to investigate such correlations, defining a probability density function (PDF) in the parameter space and providing a consistent estimate of the uncertainties. This approach is becoming more and more widespread in seismological applications, e.g., for earthquake location (Lomax *et al.* 2000), kinematic source characterization (Song & Somerville 2010) and kinematic source inversion (Piatanesi *et al.* 2007; Minson *et al.* 2013). A Bayesian approach has also been proposed for the estimation of earthquake source parameters (Garcia-Aristizabal *et al.* 2016). However, this approach is focused on prior states based on expert opinions of seismologists that require visual inspection of displacement spectra and thus it is not suitable for massive analysis of large databases. Moreover, the technique is built fixing the quality factor and then directly mixing the estimations from single spectra in a single PDF. On the one hand it neglects existing correlations among source parameters and quality factor, on the other hand final estimates are poorly constrained and show a large variability, with retrieved moment magnitude for the analyzed earthquake ranging between 5 and 9.

Here we develop a probabilistic approach using the framework defined by Tarantola (2005), based on the conjunction of states of information between data and model, to jointly retrieve source parameters and Q with uncertainties and to analyze the between-parameter correlations. Following this approach, we first search for the maximum of the a-posteriori PDF, then we compute statistical parameters through integration of the PDF. Several exploration techniques have been proposed to seek for the maximum (or minimum) of an objective, non-linear function. With the development of computational resources, more often global exploration methods are applied to inverse problems in geophysics, to limit the possibility to fall in local maxima or minima, such as the simulated annealing (Kirkpatrick *et al.* 1983; Mosegaard & Vestergaard 1991), the genetic algorithm (Goldberg & Holland 1988; Sen & Stoffa 1992; Festa & Zollo 2006), the neighborhood algorithm (Sambridge 1999; Marson-Pidgeon *et al.* 2000; Lucca *et al.* 2012) and the Monte Carlo exploration (Sambridge & Mosegaard 2002). Sometimes they are coupled with local solvers, such as the Simplex (Zollo *et al.* 2000), to speed up the convergence, reducing the number of the needed evaluations of the cost function of at least one order of magnitude (Kerner *et al.* 2008). Here we adopt the basin hopping technique (Wales & Doye 1997; Wales 2003), which couples a global exploration built on a Markov chain with a local deterministic maximization/minimization, based on a quasi-Newton algorithm. The technique improves the convergence toward the maximum or the minimum of the selected objective/cost function while maintaining a global view on the function; in addition, it is based on sole three parameters, allowing to shape the technique to the specific problem with fine tuning of the parameters. In this work we present a probabilistic approach for the characterization of the earthquake source parameters. In the first section we setup the physical quantities to be estimated and define the forward operator, built on the generalized Brune's model to be used for the inverse problem. In the next section, we summarize the main ingredients of the probabilistic framework and we adapt it to the specific problem of retrieving the source parameters. Here, we detail the strategy for the exploration and sampling of the a-posteriori PDF. In a first step we search for the maximum of the PDF and then we analyze its shape in a domain centered around it. After calculation of the marginal PDF for single parameters and for couples of parameters, we define uncertainties and correlation coefficients, and we establish a quality criterion allowing to automatically reject unconstrained solutions.

Then we perform synthetic tests to assess theoretical correlations and define the minimum bandwidth for parameter resolution. As an application, we finally estimate the source parameters for the 2016-2017 Central Italy seismic sequence.

## 2. Generalized Brune's model

In linear elasticity, the displacement produced by a point source and recorded at a given receiver is the convolution of the source time function by the Green's propagator. Thus, the displacement spectral amplitude in the frequency domain $\tilde{u}(f)$ can be factorized as $\tilde{u}(f) = \tilde{S}(f) \cdot \tilde{G}(f)$, where $f$ is the frequency, $\tilde{S}(f)$ is the modulus of the Fourier transform of the source time function and $\tilde{G}(f)$ the modulus of the Green's propagator. We separately model the far field P- and S-waves. For the source contribution, we consider the generalized Brune's model (Brune 1970):

$$\tilde{S}(M_0, f_c, \gamma; f) = \frac{M_0}{1 + \left(\dfrac{f}{f_c}\right)^{\gamma}} \qquad (1)$$

with $\tilde{S}$ depending on three parameters: the seismic moment $M_0$, the corner frequency $f_c$ and the high frequency spectral decay factor $\gamma$. The Green's function in the spectral domain can be written as (Aki & Richards 1980)

$$\tilde{G}(Q^{\prime c}; f) = K^c A^c(\mathbf{r}, \mathbf{r}_0) e^{-\pi f T^c(\mathbf{r}, \mathbf{r}_0) Q^{\prime c}} \tilde{H}(f) \qquad (2)$$

In the above formula, $K^c$ is a constant, depending on the source-receiver geometrical configuration and the elastic properties of the medium crossed by the waves, $A^c(\mathbf{r}, \mathbf{r}_0)$ and $T^c(\mathbf{r}, \mathbf{r}_0)$ are the geometrical spreading and the travel-time related to the selected wave from the source at $\mathbf{r}_0$ to the receiver at $\mathbf{r}$, respectively, and $Q^{\prime c} = \dfrac{1}{Q^c}$ is the reciprocal of the quality factor $Q^c$, which may depend on the frequency. In our modelling we assume a frequency independent quality factor, which is considered a good approximation in a broad frequency domain (Aki & Richards 1980). Finally, $\tilde{H}(f)$ is the site amplification factor and it can also depend on the frequency. In this study we neglect the contribution of the site and we assume $\tilde{H}(f) = 1$.

For a 1D layered model the analytical representation of the constant $K^c$ is (Aki & Richards 1980)

$$K^c = \frac{R^c_{\theta\varphi} F_S}{4\pi \rho(\mathbf{r}_0)^{1/2} \rho(\mathbf{r})^{1/2} c(\mathbf{r}_0)^{5/2} c(\mathbf{r})^{1/2}} \qquad (3)$$

Here $R^c_{\theta\varphi}$ is the radiation pattern contribution, depending on the phase $c$ (P or S wave), $F_S$ is the free-surface correction coefficient, $\rho$ is the density and $c$ the wave velocity. The geometrical spreading describes how the amplitude decays as a function of the distance from the source; both the geometrical spreading and the travel time can be computed using the ray theory. They simplify to $A^c(\mathbf{r},\mathbf{r}_0) = \dfrac{1}{\|\mathbf{r}-\mathbf{r}_0\|}$ and $T^c(\mathbf{r},\mathbf{r}_0) = \dfrac{\|\mathbf{r}-\mathbf{r}_0\|}{c}$ for a homogeneous medium, and the geometrical spreading is independent of the phase.

In our modelling, we assume $K^c$, $A^c(\mathbf{r},\mathbf{r}_0)$ and $T^c(\mathbf{r},\mathbf{r}_0)$ known; they are computed either in a homogeneous or in a 1D horizontally layered medium. Uncertainties in these terms contribute to the increase of the epistemic uncertainties on the source parameters and $Q^{ic}$ estimations. Both the geometrical spreading and the constant $K^c$ are scale factors for the seismic moment. They depend on the relative location of the source and the receiver and on the velocity structure crossed by the waves. However, because of the logarithmic scale of the seismic moment, their uncertainties poorly affect the estimation of the event magnitude. The travel-time appears in formula (2) through the product $Q^{ic} \cdot T^c(\mathbf{r},\mathbf{r}_0)$, where the $Q^{ic}$ factor is retrieved from the inversion of the displacement spectra. Uncertainty on $T^c(\mathbf{r},\mathbf{r}_0)$ only affects the final estimate of $Q^{ic}$ and can be completely absorbed in the inversion of the quality factor.

Because of the exponential nature of the seismic moment, we define the forward operator as the logarithm of the displacement spectral amplitude:

$$\log \tilde{u} = \log M_0 - \log\left[1 + \left(\dfrac{f}{f_c}\right)^\gamma\right] + \log \xi - \pi f T^c(\mathbf{r},\mathbf{r}_0) Q^{ic} \log e \qquad (4)$$

where $\xi = K^c A^c(\mathbf{r},\mathbf{r}_0)$; $\log \tilde{u}(\log M_0, f_c, \gamma, Q^{ic})$ depends on four unknown parameters that will be inverted analyzing the spectra obtained from seismic records. For sake of simplicity, we summarize the set of parameters to be estimated through the vector $\mathbf{m} = (\log M_0, f_c, \gamma, Q^{ic})$ and we indicate with $\mathbf{M}$ the model space, the subdomain of $R^h$, $h = 4$, which individuates the range of variability of the model parameters.

## 3. Source parameters inversion

### 3.1. Probabilistic framework for the inverse problem

Although the equation (4) provides a continuous mapping between the parameter space and the theoretical amplitude spectrum, the displacement spectra obtained from observations are sampled at a discrete, finite set of points. Let us indicate with $\mathbf{d}_{obs} = \{\log \tilde{u}_{obs}(f_k), k = 0, 1, .., n\}$ the logarithm of the discrete Fourier amplitude spectrum computed from the observed displacement; the vector $\mathbf{d}_{obs}$ belongs to the data space $\mathbf{D}$. In the above relationship, $f_k = k f_{min} = \frac{k}{T}$, where $T$ is the window length of the selected signal in time, $f_{min} = \frac{1}{T}$ the minimum frequency in the spectrum, $n$ is half of the number of samples in the time domain, and $f_n = n f_{min} = \frac{1}{2\Delta t}$ the Nyquist frequency, where $\Delta t$ is the time step of the recorded signal. It is worth to note that when using the FFT for spectral computation, the signal is padded to zero to satisfy the condition that $n$ is a power of two. To compare theoretical and observed spectra, we then compute the theoretical prediction at the same discrete set of frequencies. Let us indicate with $\mathbf{g}(\mathbf{m}) = \{\log \tilde{u}(\mathbf{m}; f_k), k = 0, .., n\}$ the discrete forward operator. The solution of the inverse problem is indeed the set of parameters $\mathbf{m}^*$, such as $\mathbf{g}(\mathbf{m}^*)$ approaches $\mathbf{d}_{obs}$ at best.

We introduce a probabilistic framework for the resolution of the inverse problem (Tarantola 2005). Let us indicate the a-priori probability density function over the model space $\mathbf{M}$, $\rho_M(\mathbf{m})$, representing the information available for the model parameters independently of the observations. Analogously, the a-priori probability density over the data space $\mathbf{D}$ is indicated as $\rho_D(\mathbf{d}, \mathbf{d}_{obs})$, representing the results of the measurement operation. Finally, we define the conditional probability density of obtaining a data vector $\mathbf{d}$ for the given set of parameters $\mathbf{m}$, $\theta(\mathbf{d}|\mathbf{m})$, that represents the information about the model prediction and its uncertainties and is thus connected to the physical theory we are using to model the observations. Combining the two states of information $\rho_M(\mathbf{m}) \cdot \rho_D(\mathbf{d}, \mathbf{d}_{obs})$ and $\theta(\mathbf{d}|\mathbf{m})$ through the conjunction operation leads to the *a-posteriori* PDF over the parameter space, which is the solution of the inverse problem for the specific observed data $\mathbf{d}_{obs}$:

$$\sigma_M(\mathbf{m}) = K \rho_M(\mathbf{m}) L(\mathbf{m}, \mathbf{d}_{obs}) \qquad (5)$$

where $L(\mathbf{m}, \mathbf{d}_{obs}) = \int_D \rho_D(\mathbf{d}, \mathbf{d}_{obs}) \theta(\mathbf{d}|\mathbf{m}) d\mathbf{d}$ is the likelihood function and $K$ is a normalization constant. Within this approach the best model $\mathbf{m}^*$ is the value that maximizes the PDF (5); however, we can also integrate $\sigma_M(\mathbf{m})$ to infer statistical indicators, such as the mean value of the distribution, the standard deviation and the correlation coefficients.

We assume that the *a priori* PDF over model space is uniform; moreover, we assume that both modelization and data uncertainties are normally distributed:

$$\theta(\mathbf{d}|\mathbf{m}) = \left((2\pi)^n \det \mathbf{C}_m\right)^{-1/2} \exp\left[-\frac{1}{2}(\mathbf{g}(\mathbf{m}) - \mathbf{d})^T \mathbf{C}_m^{-1}(\mathbf{g}(\mathbf{m}) - \mathbf{d})\right] \quad \text{and}$$

$\rho_D(\mathbf{d}, \mathbf{d}_{obs}) = \left((2\pi)^n \det \mathbf{C}_d\right)^{-1/2} \exp\left[-\frac{1}{2}(\mathbf{d} - \mathbf{d}_{obs})^T \mathbf{C}_d^{-1}(\mathbf{d} - \mathbf{d}_{obs})\right]$, with $\mathbf{C}_m$ the covariance matrix related to modelization uncertainties and $\mathbf{C}_d$ the covariance matrix related to measurement uncertainties. Under this hypothesis the likelihood writes (Tarantola 2005, pp. 35-36;202-203):

$$L(\mathbf{m}, \mathbf{d}_{obs}) = \exp(-S(\mathbf{m}, \mathbf{d}_{obs})) \qquad (6)$$

where $S(\mathbf{m}, \mathbf{d}_{obs}) \equiv \frac{1}{2}(\mathbf{g}(\mathbf{m}) - \mathbf{d}_{obs})^T \mathbf{C}_D^{-1}(\mathbf{g}(\mathbf{m}) - \mathbf{d}_{obs})$ is the cost function, and $\mathbf{C}_D = \mathbf{C}_d + \mathbf{C}_m$ is the total covariance matrix. It is worth to note that this relation is general, regardless the linearity of the $\mathbf{g}(\mathbf{m})$ operator.

We assume that the covariance matrix is diagonal and it has the following form $\mathbf{C}_D = \mathbf{1} \cdot MSE$, where $MSE = \sum_{i=1}^{n} \frac{\left[(d_{obs})_i - g_i(\mathbf{m}^*)\right]^2}{n - h}$. Under this assumption maximization of the likelihood function also corresponds to the minimization of the $L^2$ distance between data and predictions,

$S'(\mathbf{m}, \mathbf{d}_{obs}) = (\mathbf{g(m)} - \mathbf{d}_{obs})^T (\mathbf{g(m)} - \mathbf{d}_{obs})$, which does not depend on the *MSE* and can be computed independently of the knowledge of the solution $\mathbf{m}^*$. In the following application on real data we found *a-posteriori* that the MSE is in the range 0.01-0.03. The data uncertainties are associated to the signal to noise ratio (SNR). When selecting data with a SNR > 10 on average in the selected frequency range, the contribution of the noise to the *MSE* is at least one order of magnitude smaller. Thus, the *MSE* is dominated by the modelization uncertainties within this range of SNR values and it is retrieved to be independent of the specific SNR value.

The solution of the inverse problem is computed in two steps : we first compute the minimum of the cost function $S'(\mathbf{m}, \mathbf{d}_{obs})$, using the basin hopping technique, as described in the next section, then we evaluate the *MSE*, which is used for the estimation of the *a-posteriori* PDF $\sigma_M(\mathbf{m})$. Finally, the estimation of the uncertainties requires the integration of $\sigma_M(\mathbf{m})$.

If the forward operator $\mathbf{g(m)}$ is linear, $S(\mathbf{m}, \mathbf{d}_{obs})$ is quadratic and $\sigma_M(\mathbf{m})$ is normal; the more nonlinear $\mathbf{g(m)}$, the farther $\sigma_M(\mathbf{m})$ from a Gaussian PDF. However, though strongly non-linear, the forward operator $\mathbf{g(m)}$ can be linearized in the vicinity of the best model, in a subdomain $\mathbf{M}^*$ centered around $\mathbf{m}^*$. If the value of $\sigma_M(\mathbf{m})$ is enough small outside $\mathbf{M}^*$, to not significantly contribute to the marginal PDFs related to the single parameters, we can extract the mean and the variance for each parameter, and the correlation coefficients for all couples of parameters, limiting the exploration to the domain $\mathbf{M}^*$. We define $\mathbf{M}^*$ as the hypercube $\mathbf{M}^* = I_{m_1} \times .. \times I_{m_h}$, where $I_{m_i}$ is a 1D interval containing the value $m_i^*$. Let us define the marginal PDF for the parameter $m_i$ as $\hat{\sigma}_M(m_i) = \int_{\mathbf{M}_i^*} \sigma_M(\mathbf{m}) d\mathbf{m}$, and the marginal PDF for the couple $(m_i, m_j)$ as $\breve{\sigma}_M(m_i, m_j) = \int_{\mathbf{M}_{ij}^*} \sigma_M(\mathbf{m}) d\mathbf{m}$, where $\mathbf{M}_i^* = I_{m_1} \times .. \times I_{m_i-1} \times I_{m_i+1} \times .. \times I_{m_h}$ is the hypercube built accounting for all the parameters except $m_i$ and $\mathbf{M}_{ij}^*$ the hypercube built excluding the parameters $m_i$ and $m_j$. Mean value, variance and correlation are finally computed as:

$$\mu_i = \int_{\mathbf{M}^*\backslash\mathbf{M}_i^*} m_i \cdot \hat{\sigma}_M(m_i) dm_i$$

$$\sigma_i^2 = \int_{\mathbf{M}^*\backslash\mathbf{M}_i^*} (m_i - \mu_i)^2 \cdot \hat{\sigma}_M(m_i) dm_i \quad (7)$$

$$\mathrm{cov}_{i,j} = \int_{\mathbf{M}^*\backslash\mathbf{M}_{ij}^*} (m_i - \mu_i)(m_j - \mu_j) \cdot \breve{\sigma}_M(m_i, m_j) dm_i dm_j$$

### 3.2. The basin hopping algorithm for the search of the global minimum

The search for the minimum of the cost function $S'(\mathbf{m}, \mathbf{d}_{obs})$ is performed through the global optimization technique of the basin hopping. It uses a random sampling of the model space, based on a Markov chain with a transition probability given by the Metropolis criterion.

Here we shortly summarize the searching strategy of the technique. If after $j$ iterations the exploration has reached the point $\mathbf{m}_j$, at the $(j+1)$-th iteration a random perturbation of the coordinates is performed, moving the model in the point $\mathbf{m}_{j+1}^{(0)}$; this latter is considered as the starting point for a local minimization, which brings the exploration in the point $\mathbf{m}_{j+1}$. The minimization is performed using the Broyden–Fletcher–Goldfarb–Shanno algorithm (Fletcher 1987). The point $\mathbf{m}_{j+1}$ is then compared with $\mathbf{m}_j$. If the cost function at the end of the $(j+1)$-th iteration is smaller than the cost function at the end of the $j$-th iteration, i.e. $S'(\mathbf{m}_{j+1}, \mathbf{d}_{obs}) < S'(\mathbf{m}_j, \mathbf{d}_{obs})$, the transition from $\mathbf{m}_j$ to $\mathbf{m}_{j+1}$ is accepted, else it is accepted with a probability $P_{trans}(\mathbf{m}_j, \mathbf{m}_{j+1})$ given by the Metropolis criterion: $P_{trans} = \exp\left(-\dfrac{S'(\mathbf{m}_{j+1}, \mathbf{d}_{obs}) - S'(\mathbf{m}_j, \mathbf{d}_{obs})}{T}\right)$, where the temperature $T$ of the Metropolis scheme is fixed all along the exploration.

It is worth to note that the solution at the end of each iteration $\mathbf{m}_j$ comes from a local minimization process, speeding up the search for the final solution, while maintaining a constant temperature facilitates the hopping out of cost function basins which contain local minima.

The method is based on sole three parameters, allowing for simple tuning. The first one is related to the initial modulus of the random perturbation. The perturbation is assumed to be the same fraction $\beta$ of the range of variability for all the parameters. If the exploration of the parameter $m_i$ is constrained in the domain $(m_{i,\min}, m_{i,\max})$, the size of the initial perturbation is thus $\Delta m_{i,0} = \beta(m_{i,\max} - m_{i,\min})$. It is worth to note that the magnitude of the perturbation dynamically changes during the exploration. It is based on the fraction $a_r$ of the accepted transitions from the point $\mathbf{m}_j$ to $\mathbf{m}_{j+1}$ over a fixed number of iterations. Here we evaluate $a_r$ every 50 iterations; if $a_r > 0.5$, $\beta$ is increased dividing the previous value by 0.9; if $a_r < 0.5$, $\beta$ is decreased multiplying the previous value by 0.9. This condition allows to explore farther and farther regions when the solution does not move from the same location in the model space for many iterations.

The second parameter is the temperature $T$ of the Metropolis criterion, which is chosen by balancing the ability to converge toward the final solution and the possibility to escape from a local minimum. Its magnitude order should be comparable with the average difference between the local minima, and thus it requires preliminary investigation. Finally, the exploration stops when the maximum number of iterations $n_{iter}$ is reached.

The tuning of the parameters is problem dependent. From synthetic tests on theoretical spectra, we obtained convergence to the global minimum with $\beta = 0.1$, $n_{iter} = 10000$ and $T$ of the order of the unity.

### 3.3. Definition of M* for uncertainty computation

The use of the joint PDF allows not only to seek for the best solution, but also to compute the uncertainties related to the best model, via integration of $\sigma_M(\mathbf{m})$. We cannot use the parameter space exploration from the basin hopping technique, because it does not rely on a Monte Carlo sampling and thus, convergence of integrals is not guaranteed when increasing the iteration number. On the other hand, a complete description of $\sigma_M(\mathbf{m})$ in the whole parameter space is computationally expensive and, in many cases, unnecessary, since this function very often rapidly decreases to zero when moving away from the maximum. For this reason, we limit the computation of the joint PDF in the hypercube $\mathbf{M}^*$ centered on the best fit model $\mathbf{m}^*$. For the definition of the hypercube we explore the 1D conditional distributions

$$\tilde{\sigma}_{M_i}(m_i) = \sigma_M(m_1^*,...,m_i,...,m_h^*) \quad i \in \{1,...,h\} \tag{8}$$

where all parameters are fixed to the value that they have in the global maximum of the PDF while the parameter $m_i$ can vary. We then define the interval $I_i^* = [m_{i\text{down}}, m_{i\text{up}}]$, containing the value $m_i^*$, such that $\tilde{\sigma}_{M_i}(m_{i\text{down}}) \approx \tilde{\sigma}_{M_i}(m_{i\text{up}}) \approx 0.05 \tilde{\sigma}_{M_i}(m_i^*)$. In the case in which the conditional PDF can be described by a Gaussian function, the interval $I_i^*$ is symmetric around $m_i^*$ and its length is four times the standard deviation of the Gaussian function. Since the marginal PDF has usually a larger standard deviation than the conditional PDF because of the correlations among parameters, we consider the enlarged interval $I_i = [\bar{m}_{i\text{down}}, \bar{m}_{i\text{up}}]$. $\bar{m}_{i\text{down}} = \max\left[m_{i\min}, (1-\lambda)m_i^* + \lambda m_{i\text{down}}\right]$ being $\lambda = 5.0$ the scaling factor between the marginal and the conditional PDF standard deviations; analogously $\bar{m}_{i\text{up}} = \min\left[m_{i\max}, (1-\lambda)m_i^* + \lambda m_{i\text{up}}\right]$. Finally, the domain $\mathbf{M}^*$ is obtained by tensorization: $\mathbf{M}^* = I_1 \times .. \times I_h$.

### 3.4. The quality of the solution

We can finally check *a-posteriori* the assumption of Gaussian uncertainties, evaluating the quality of the retrieved marginal PDFs $\hat{\sigma}_M(m_i)$ in terms of similarity with a normal distribution. As similarity criterion, we adopt the normalized cross-correlation function:

$$cc(\tau) = \int_{\mathbf{M}_i^*} \hat{\sigma}_M(m_i)\sigma_{\exp}(m_i+\tau)dm_i \qquad (9)$$

where $\sigma_{\exp}(m_i) \equiv N(m_i;\mu_i,\sigma_i^2)$ is the expected, normal distribution having median $\mu_i$ and variance $\sigma_i^2$. We selected a quality threshold $\theta_\sigma$; if the zero-lag correlation $cc(0) \geq \theta_\sigma$, the solution is accepted. In our analysis we chose $\theta_\sigma = 0.95$.

## 3.5. Dataset features and final results expression

In the previous sections, we described how to retrieve source parameters from the inversion of a single spectrum. However, an earthquake is recorded at several stations, usually on the three components of a seismic instrument. For a single station, we invert for one spectrum for the P wave, obtained from the vertical component, and one spectrum for the S wave, obtained as the geometrical mean of the two spectra, computed on the horizontal components (Fletcher *et al.* 1984). When combining information from diverse stations we should be aware of the fact that each station provides a different image of the earthquake source, depending on the directivity, the radiation pattern and propagation effects. For that reason, we cannot consider each spectrum as a repeated measure of the same source parameters. After obtaining an estimation of the source parameter $(\mu_i)_k$ with uncertainty $(\sigma_i)_k$ from the inversion of a single spectrum at the *k*-th station, the final estimation of this parameter is given by the weighted mean

$$\bar{\mu}_i = \frac{\sum_{k=1}^{K}(\mu_i)_k (w_i)_k}{\sum_{k=1}^{K}(w_i)_k} \qquad (10)$$

and the weighted uncertainty is given by

$$\bar{\sigma}_i = \sqrt{\frac{1}{\sum_{k=1}^{K}(w_i)_k}} \qquad (11)$$

where $(w_i)_k = \dfrac{1}{(\sigma_i^2)_k}$ and $K$ is the total number of stations contributing to the mean.

## 4. Synthetic tests

To assess the method performances, we realize several synthetic tests, where we completely control the input signals to be inverted for the retrieval of the source parameters. Synthetic signals are generated assuming a perfect knowledge of the propagation (geometrical spreading, elastic properties and travel time of the wave) which does not affect the final estimates and related uncertainties. The signal is generated according to the generalized Brune's model of formula (2). For the tests we used the following parameters $\log M_0 = 10$, $f_c = 10\ Hz$, $\gamma = 2$. The signal is then polluted by noise with an assigned SNR value. It is worth to note that the SNR is defined in the time domain. Here the displacement $u$ is the sum of the signal $s(t)$ and the noise $n(t)$ : $u(t) = s(t) + n(t)$. The displacement amplitude spectrum can be written as (see Appendix A) $\log(\tilde{u}) \approx \log(\tilde{s}) + \dfrac{\tilde{n}}{\tilde{s}} \cos(\varphi_S - \varphi_N)$, where $\tilde{s}$ and $\tilde{n}$ are the amplitude spectra and $\varphi_S$ and $\varphi_N$ the phase spectra of the signal and the noise, respectively; they are all functions of the frequency. The ratio $\dfrac{\tilde{n}}{\tilde{s}}$ scales as $\dfrac{1}{SNR}$ and at low frequencies, for a flat noise spectrum the ratio $\dfrac{\tilde{n}}{\tilde{s}} \approx \dfrac{1}{SNR}$. However, the noise spectrum is usually not flat in the displacement, but it decreases as a function of the frequency as well as the source spectrum does, eventually with different slopes. In addition, the noise spectrum is not uncoherent, but it presents peaks and holes related to ambient and site effects.

In these tests we impose the following perturbation on the Brune's spectrum $\tilde{u}_{Brune}$ to account for the noise effect in the displacement:

$$\log \tilde{u} = \log \tilde{u}_{Brune} + \dfrac{1}{SNR} \sin\left(\cdot \dfrac{2\pi f}{f_N}\right) \cdot (1+\eta) \qquad (12)$$

The noise has a coherent contribution having a sinusoidal shape with amplitude equal to the reciprocal of the signal to noise ratio, modulated by a random contribution. Here we have $f_N = 1\,Hz$ and $\eta$ is a random variable over the interval $[-0.5, 0.5]$.

In this study we present four synthetic tests. We first analyze the case of a very large SNR value, to check the method, to understand the relation between the SNR and the final uncertainties in the inverted parameters and to estimate the intrinsic correlations among parameters. We thus use a more realistic SNR value, which is also representative of the SNR values observed in the case study discussed in the next section, to investigate the effect of the SNR on the final solutions.

In many applications, however, the ratio between the spectral level of the signal and the spectral level of the noise is not constant across the whole frequency range; very often the noise has the same level as the signal at both low and high frequencies, leaving a limited bandwidth around the corner frequency for the analysis. We then explore how large the theoretical bandwidth should be to have an unbiased estimation of the source parameters and of the quality factor.

As a last test, we analyze potential biases on the inverted parameters due to an effective frequency dependent quality factor, when performing the inversion using a constant Q value.

### 4.1. Large signal to noise ratio

As a first example we want to show the reliability of the method on a signal poorly affected by the noise. This example will also enable us to understand how the uncertainties are computed and what are the intrinsic correlations between couples of parameters. In this test we further set $Q = 100$ and SNR = 100. The spectrum has been inverted in the frequency band [0.1 $Hz$ -100 $Hz$].

Within this high value of the signal to noise ratio, the solution almost perfectly fits the initial spectrum (Figure 1), indicating the effective convergence of the method.

The 1-D marginal PDFs are Gaussian-like distributions, centered on the values imposed to generate the synthetic spectrum (Figure 2), pointing out also the uniqueness of the solution. We have the following estimates for the source parameters: $\log M_0 = 10.000 \pm 0.004$, $f_c = 9.99 \pm 0.09\ Hz$ and $\gamma = 1.999 \pm 0.015$, and the final estimate for the quality factor is $Q = 100.00 \pm 0.05$, whose uncertainty has been obtained by propagating the error on $Q'$. For this case, we have very small uncertainties on the seismic moment and the quality factor (< 0.1%), while the uncertainty is about at 1% on both $f_c$ and $\gamma$.

In Figure 3 we represent the 2-D marginal PDFs as heatmaps. Since the data uncertainty is negligible in this case, the maps represent the intrinsic correlation between couples of parameters. This correlation is due to the modelization uncertainty and cannot be reduced. The absolute value of all the correlation coefficients is above 0.6; larger (anti-)correlations can be found between $f_c$ and $\log M_0$, $\gamma$ and $Q'$, with values close to -1. These large correlations justify the use of the factor $\lambda$, to define the exploration interval for the computation of the marginal PDFs from the conditional PDFs (Section 3.3). Also, large values for the correlation coefficients indicate that the use of robust *a-priori* information can strongly reduce the uncertainty on the final parameter estimates.

### 4.2. Signal to noise ratio SNR = 5

We perform a test similar to the previous example with $Q = 100$ and a smaller signal to noise ratio (SNR = 5), this value being representative of the average SNR value in the analyzed dataset. The spectrum is inverted in the frequency band [0.1 *Hz* -100 *Hz*]. In Figure 4 we represent the comparison between the retrieved solution (red curve) and the synthetic spectrum (blue curve). Despite the decrease of the SNR value, we still have a good estimate of the parameters, but uncertainties are larger. The final estimates are $\log M_0 = 10.01 \pm 0.08$, $f_c = 9.7 \pm 1.7\ Hz$, $\gamma = 1.9 \pm 0.3$ and $Q = 99.7 \pm 1.1$. Although the smallest uncertainties are still retrieved for the seismic moment and the quality factor, now they have increased to 0.8% and 1.1% respectively. The percentage error is increased to 18% on $f_c$ and to 16% on $\gamma$. These values can be considered as a lower bound for the relative uncertainties in the real application shown in the next section, where also the epistemic uncertainty on the Green propagator can affect the results. In Figure 4 we compare the marginal distribution for this case with the case of SNR = 100 for the corner frequency, on the same scale. The distribution still maintains its Gaussian shape, but its width is significantly increased. This also occurs for the other three marginal distributions related to $\log M_0$, $\gamma$ and $Q'$. It is worth to note that, on the contrary, correlation coefficients do not significantly change (Figure 5): thus, correlation between couples of parameters is mostly due to the modelization uncertainty, and the contribution from data is negligible.

### 4.3. Frequency bandwidth for the inversion

The resolution of the source parameters and of the quality factor depends on the frequency bandwidth available for the inversion. In the previous examples the bandwidth for the inversion is enough large, with two decades before the corner frequency and one decade after it, to allow the resolution of the parameters. We want to investigate how the resolution degrades when we shrink the bandwidth around the corner frequency. In this test we use a value of SNR = 5 and $Q = 100$. We analyze the reduction of the bandwidth with three different approaches: we reduce the band at frequencies larger than $f_c$ (case 1), smaller than $f_c$ (case 2) and symmetrically around $f_c$ (case 3). The results are shown in Figure 6, panels A, B and C respectively. In Figure 6 we represent the relative difference between the expected and the retrieved values, for all the parameters. In all cases, when we reduce the band, the uncertainty increases and eventually a bias in the estimation of two or more parameters can emerge because of the unresolved correlations. In Figure 6A (case 1) we can observe that the parameters are well resolved down to a frequency band as large as 0.4 decade. The seismic moment and the quality factor are almost unsensitive to the reduction of the high-frequency band, while the uncertainty in both $f_c$ and $\gamma$ significantly increases as the bandwidth decreases. The correlations start to be poorly resolved at a width of 0.4 decade beyond the corner frequency. The mean value of all the parameters is less sensitive to the reduction of the bandwidth at the left of the corner frequency (case 2, Figure 6B). Still larger errors arise from the correlation between $f_c$ and $\gamma$ but with mean values well constrained down to a 0.1 decade. When a symmetric restriction is performed the quality of the solution is controlled by the high-frequency region and again we need a bandwidth of 0.4 decade to have proper resolution on the parameters (case 3, Figure 6C). Finally, in Figure 7 we show the heatmaps of the correlation between $Q'$ and $f_c$ for a symmetric bandwidth size of 0.3 decade and 0.4 decade around the corner frequency. When reducing the bandwidth, we see a migration of the maximum of the marginal PDF toward the upper limit of the explored frequency band, with the smoother decay of the spectrum around $f_c$ being instead explained by a slightly lower $Q$ value. The test thus indicates that we need at least 0.4 decade around the corner frequency to resolve the source parameters. Eventually this band can be reduced at the left of the corner frequency. The same result holds also when considering a noise linearly decreasing with the frequency, which is representative of a noise with flat spectrum in velocity.

It is worth to note that the accuracy in the quality factor estimate depends on the value of $Q$ itself: the smaller the $Q$ the more relevant its effect on the spectrum. The uncertainty on the $Q$ estimate increases as $Q$ increases. If we consider the solutions for $Q = 100$ and $Q = 800$ obtained from a theoretical spectrum inverted in the same frequency band, symmetric around $f_c$ with 0.4 decade available on the two sides of the corner frequency, the percentage error moves from 5 % to 37 %, with the estimates for the quality factor of $Q = 98 \pm 5$ and $Q = 680 \pm 250$, respectively. However, the change in the $Q$ value does not affect significantly the accuracy and the quality of the solutions for the other parameters.

### 4.4. Inversion from a model with a frequency dependent Q

We finally analyze possible biases due to the use of a simplified, frequency independent attenuation model. We generated a synthetic spectrum using a frequency dependent quality factor according to the formula $Q(f) = Q_0 f^\varepsilon$ (e.g. Aki 1980). We used the following values for the parameters : $Q_0 = 300$, $\varepsilon = 0.3$ and SNR=5; this corresponds to a quality factor varying from Q = 150 (f = 0.1 Hz) to Q = 1200 (f = 100 Hz). Observations in different regions of the world (Takahashi 2012; Atkinson 1995) are consistent with this model. The inversion is then performed assuming a constant, frequency independent $Q$ value. The retrieved solution well fits the synthetic spectrum, as shown in the upper panel of Figure 8. When looking at the marginal PDF for the source parameters (Figure 8) we see that the estimates for the source parameters are consistent with the input values within uncertainties. These latter are comparable with the ones obtained for the same SNR value using the input spectrum generated with a constant $Q$ value (Section 4.2). However, the estimate for $Q$ is $Q = 1600 \pm 300$ and the associated PDF is close to zero in the range of values assumed by $Q(f)$.

Thus, in an attenuating structure in which the quality factor changes with the frequency, an inversion performed assuming a constant $Q$ value will still provide a good estimation of the source parameters. However, we should keep in mind that the retrieved $Q$ value does not necessarily correspond to the average value of $Q$ with frequency and requires further independent check.

## 5. Application to the 2016-2017 Central Italy sequence

### 5.1. Data

A major earthquake sequence interested the Central Italy region from August, 2016 to January, 2017. The first event of the sequence, the $M_w = 6.0$ Amatrice earthquake occurred on August 24, 2016 with epicenter in the village of Accumoli and caused 298 casualties and more than 15000 displaced persons. The largest earthquake of the sequence – the $M_w = 6.5$, Norcia earthquake - occurred on October 30, 2016 with epicenter in the village of Norcia. It generated large slip at the surface, with a maximum observed dislocation of about 2 *m*.

We computed the source parameters for the major events of this sequence. The dataset consists of accelerometric records for all the events of the sequence with $M_L \geq 4.0$ (56 events), recorded by stations within 100 *km* from the epicenter. In Figure 9 we represent the location of the events and the stations: we have at maximum 62 stations per event and the hypocentral distance ranges between 9 *km* and 100 *km*. The total number of records is 1909.

The sampling frequency is 100 *Hz*, 125 *Hz* and 200 *Hz*, depending on the station.

The waveforms were downloaded from ESM (Engineering Strong-Motion database) (Luzi *et al.* 2016); the metadata were acquired from INGV bulletin (ISIDe working group 2016).

### 5.2. Processing

For each record, the definition of the S-wave (signal) duration $\Delta t$ was based on the expected ground motion duration (Trifunac & Brady 1975):

$$\Delta t = \frac{0.02 \cdot \exp(0.74 \cdot M_L) + 0.3 \cdot \delta_H}{a} \tag{13}$$

where $\delta_H$ is the hypocentral distance and $a$ is a factor introduced to rescale the ground motion duration to the S-wave duration; we chose $a = 4$ for the largest events in the dataset ($M_L \geq 5.9$) and $a = 2$ for the remaining events. The S-wave time-window $\Delta T_S$ was defined using the theoretical S-wave arrival time $T_S$ obtained from the 1-D velocity model of Chiarabba (2009); the time window starts before $T_S$ such that the selected window is $\Delta T_S = [T_S - 0.1 \cdot \Delta t, T_S + 0.9 \cdot \Delta t]$.

A noise time-window $\Delta T_N$ of the same duration $\Delta t$ was selected before the event origin time $T_0$ as $\Delta T_N = [T_0 - \Delta t, T_0]$. With this choice, we do not include the P-wave within the noise window. Since the direct P wave does not usually perturb the S waveform, but eventually only the P coda may affect it, using the P wave as noise to be compared to the S signal artificially amplifies the noise contribution.

On both signal and noise waveforms we removed constant and linear trends and applied a Hann-function tapering on the first and last 5 % of the data. FFT was computed for pre-processed signal and noise; multiplication by $1/(4\pi^2 f^2)$ yields displacement amplitude spectra. Each spectrum has been finally smoothed with a 5-points moving average filter.

When analyzing a single spectrum, we computed the signal-to-noise ratio for each point in the frequency domain; starting from the center of the domain, we extracted the closest points at the left and at the right of $f_c$ with SNR = 1.25. These points become the bounds of the frequency domain used for the fit. As a result, this allows us to rule out the regions of the frequency domain where the noise is comparable to the signal, and to reject records having no information about the earthquake. An example of spectrum passing this first checkpoint is shown in Figure 10, where the region in the frequency domain selected for the fit is evidenced by two arrows. In the same Figure, we also represent the noise spectra with a gray line. The SNR value is larger than 10 in a large portion of the domain selected for the fit. For these data we compute the a-posteriori PDF, by first evaluating the maximum of the likelihood function and then the marginal a-posteriori PDF on the single parameters. In Figure 10 we superimpose to the observed data the theoretical spectrum computed using the solutions obtained from the PDFs (red curve). When analyzing the marginal a-posteriori PDF for the corner frequency, we see that it has a long tail, up to the maximum frequency of the data, indicating that this parameter is unconstrained. Thus, the solution for the source parameters from this spectrum cannot be considered. When comparing the shape of this PDF with a Gaussian function, we found a degree of similarity of 0.93, according to the criterion defined in Section 3.4. With a quality threshold of 0.95 this solution is automatically discarded. Applying the SNR and the similarity criteria we ruled out 701 solutions (37 % of the available records). The final estimates are based on 1208 solutions, and we have from 6 to 37 solutions per event in the dataset.

### 5.3. Results

We investigated the earthquake source properties for the main events of the Central Italy 2016-2017 sequence. In Figure 11 we show an example of S waveform in time domain and the related displacement amplitude spectrum recorded at the station FIAM for the $M_L = 4.0$ event occurred the day 24-08-2016 at 23:22:05 (UTC). Superimposed to the observed spectrum (blue curve), we also plot the noise spectrum (gray curve) and the theoretical spectrum (red curve), obtained using the mean estimate of the parameters. It is worth to note that the final solution well describes the observed spectrum. In Figure 12, we show the 1-D marginal PDFs for all the parameters which exhibit a peaked, Gaussian-like behavior. For those functions, the similarity ranges between 0.995 and 1. These solutions are thus accepted according to the pre-defined threshold.

We plot 2-D marginal PDFs as heatmaps in Figure 13. We clearly recognize the correlations among the parameters and they have almost the same shape as the heatmaps computed for the synthetic waveforms with large SNR (Figure 3). Thus, also in the application, the correlation is mainly governed by the modelization uncertainties. This shape can be observed for most of the data providing source parameters for this case study.

We finally represent in Figure 14 the solutions for all the events, where we plot the corner frequency as a function of the seismic moment. We observe on average that the standard scaling $M_0 \propto \frac{1}{f_c^3}$ holds (Aki 1967) with an average static stress drop $\Delta\sigma = 2.1 \pm 0.3\, MPa$, although we have a large variability in the corner frequency estimates for events with similar seismic moment. Specifically, for events with moment magnitude $M_W$ between 4 and 5 we report a variability in the corner frequency of a factor 5, with the stress drop jumping from few hundreds $kPa$ to 10 $MPa$. However, the majority of the events has a stress drop close to the average values within uncertainties. As the magnitude increases the stress drop on average increases. For the Norcia event ($M_W = 6.4 \pm 0.1$), we have a corner frequency $f_c = 0.15 \pm 0.03\, Hz$ and a high-frequency decay slope $\gamma = 2.14 \pm 0.08$. The estimated source radius is $r = 8.3 \pm 1.8\, km$ and the static stress drop is $\Delta\sigma = 4 \pm 3\, MPa$. For the Amatrice event, we have a moment magnitude of $M_W = 6.11 \pm 0.07$, a corner frequency $f_c = 0.27 \pm 0.04\ Hz$ and a high-frequency decay slope $\gamma = 2.05 \pm 0.08$. The estimated source radius is $r = 4.5 \pm 0.7\ km$, the static stress drop is $\Delta\sigma = 9 \pm 4\ MPa$. Finally, for the Visso earthquake, we have a moment magnitude of $M_W = 5.93 \pm 0.05$, a corner frequency $f_c = 0.23 \pm 0.03\ Hz$ and a high-frequency decay slope $\gamma = 1.94 \pm 0.04$. The estimated source radius is $r = 5.3 \pm 0.6\ km$, the static stress drop is $\Delta\sigma = 3.0 \pm 1.0\ MPa$.

Results for $\gamma$ and $Q$ are shown in Figure 15. The $\gamma$ distribution has a median value equal to 2.1, with the 60 % of events exhibiting a $\gamma$ value between 1.9. and 2.3 as expected from the standard Brune's model (Brune 1970). The $Q$ distribution has a mean value of 440, and a standard deviation of 290; although it is affected by a large error, this can be interpreted as a mean value for the anelastic attenuation factor of the whole explored region. It is consistent with the estimate of Bindi *et al.* (2004). They found an average S wave anelastic attenuation factor of 318 for the Central Italy region; it is worth to note that they used a different Green's function with a frequency-dependent Q factor and a constant Q value was found only for frequencies above 8 $Hz$.

### 5.4. Discussion

The average stress drop obtained in this study is similar to what retrieved for the two major seismic sequences, that have interested the Central Italy region (Umbria-Marche 1997-1998, L'Aquila 2009) in the last two decades. Bindi *et al.* (2004) found an average stress drop of $2 \pm 1\ MPa$ for the Umbria-Marche sequence; for L'Aquila sequence Pacor *et al.* (2015) showed a stress drop variability that spans two orders of magnitude, ranging in the interval $(0.1 - 25)\ MPa$, with an average value of $2.6\ MPa$, and a higher value - $10\ MPa$ - for the largest event $(M_W = 6.3)$. They also observed a stress drop increase from 1 to 10 $MPa$ with the moment magnitude ranging from 3 to 5.8. Del Gaudio *et al*. (2015) reanalyzed the source parameters for some events of the L'Aquila sequence, to extract appropriate empirical Green's functions for numerical simulations. They also found a self-similarity in the selected dataset, including events with magnitude ranging between 3.5 and 6.3, with an average stress drop of 3 MPa.

We report a larger stress drop for the largest magnitude events in the sequence, as already inferred by Bindi *et al*. (2018a). Also, the retrieved stress drop values are comparable with the ones from Bindi *et al*. (2018a). Instead, they are significantly lower than the ones reported by Picozzi *et al.* (2017) and computed from the ratio between radiated energy and seismic moment. For the same seismic moment, these values would require a significantly lower corner frequency, which has been not observed in our data.

In addition, a larger stress drop for the main events in this sequence has been also reported for other sequences in Central Italy (Bindi *et al.* 2017; Bindi *et al.* 2018b), although the retrieved values are slightly smaller as compared to the values obtained in literature (Bindi *et al.* 2018b).

Kinematic inversions of the major events in this sequence revealed that most of the slip has been produced in asperities whose size is significantly smaller than the fault size (Tinti *et al.* 2016; Liu *et al.* 2017; Chiaraluce *et al.* 2017). Specifically, in the case of the Amatrice earthquake, Chiaraluce *et al.* (2017) retrieved a small size slip patch ($3x3\ Km^2$) with a maximum slip larger than 1 *m*, just up-dip of the hypocenter and a secondary slip patch of about $5x5\ Km^2$, with an average slip of about 50 *cm*. On the rest of the fault ($\approx 20x15\ Km^2$) the slip level is smaller than 30 *cm*. Also, for the M 6.5 Norcia and the M 5.9 Visso earthquakes the slip is concentrated in few patches whose size is smaller than the rupture area expected for events of such a magnitude (Wells & Coppersmith 1994). Hence, larger magnitude events have been mainly generated by asperities where the stress is concentrated and thus exhibiting a large yield strength, within a fault system with an average lower level of frictional strength.

We computed the Savage-Wood seismic efficiency $\eta_{SW}$ (Savage & Wood 1971) as the ratio between the apparent stress $\tau_a$ and the stress drop. The apparent stress is computed from the radiated energy $E_R$ as (e.g. Wyss 1979):

$$\tau_a = \mu \frac{E_R}{M_0} \qquad (14)$$

The radiated energy is the sum of P and S wave contributions: $E_R = E_R^{(P)} + E_R^{(S)}$. The S-wave radiated energy can be obtained for each station-event couple from the theoretical source spectrum $\tilde{S}(f)$, built from the retrieved source parameters according to the formula (Boatwright & Flecther 1984):

$$E_R^{(S)} = \frac{1}{\rho c_S^5} \int_0^\infty f^2 \tilde{S}^2(f) df \qquad (15)$$

Finally, we use the relationship $E_R^{(P)} = \frac{3 c_S^5}{2 c_P^5 \zeta^3} E_R^{(S)}$ (Boatwright & Flecther 1984), where $\zeta$ is the ratio between the P and the S corner frequencies. We fixed $\zeta$ such that the ratio $E_R^S / E_R^P = 13.7$ (Boatwright & Fletcher 1984), and we found that the seismic efficiency only depends on $\gamma$. In our study, the high-frequency scaling $\gamma$ follows on average a $f^{-2}$ decay, as proposed in the classical models of circular ruptures (Madariaga 1976), leading to an almost constant seismic efficiency with seismic moment (Figure 16), which ranges between $10^{14.2}$ $Nm$ and $10^{18.8}$ $Nm$. Such low values of seismic efficiency are usually retrieved for earthquakes (Beeler *et al.* 2003; Zollo *et al.* 2014) and indicate an overshoot in the static final stress and compared to the dynamic frictional level. This can be related to a dynamic overshoot, indicating a frictional weakening of the interface, while the rupture is propagating on the fault, or large energy required to create new rupture surface (high fracture efficiency). Higher values of $\eta_{SW}$ are characterized by large errors. This behavior is expected because a large seismic efficiency is retrieved for $\gamma$ values close to 1.5. Since we have convergence in the estimation of the radiated energy (15) for $\gamma > 1.5$, nearby the convergence limit of $\gamma$ the variability in $E_R$ is large, thus generating a large uncertainty in the $\eta_{SW}$ estimation.

In this study, we neglected the contribution of the site amplification, increasing the epistemic uncertainty in the estimation of the source parameters and the quality factor. When the shallow structure is known, an explicit representation of the site contribution can be provided, assuming for instance a near-vertical propagation of seismic waves below the station (site transfer functions, e.g Lemo & Chavez-Garcia 1993). This contribution can be directly included in the expression of the Green function of formula (2). In most cases, however, the site amplification is not known a-priori in the broad frequency domain used for the modelling of the source properties and needs to be estimated at the same time of the source parameters. A first order correction can be introduced in the method analyzing the station residuals. Under the assumption that the mean value of the parameters $\bar{\mu} = (\bar{\mu}_1, \bar{\mu}_2, \bar{\mu}_3, \bar{\mu}_4)$ computed from eq. (10) is a good estimate of the effective source parameters and quality factor, the residual at the *k-th* station $R_k(f) = \log \tilde{u}_{obs,k}(f) - \log \tilde{u}_k(\bar{\mu}, f)$ provides a good estimation for the site amplification factor at that location.

Since diverse effects may contribute to the station residuals per single event, such as ambient noise or inaccurate corrections for large scale propagation, we need to collect the residuals for a large number of events recorded at the same station to have a good estimation of the site amplification. Thus, this contribution can be added to the forward operator of eq. (4), to account for the site effect. Under the assumption that uncertainties on the site amplification function are independent of the frequency, we can apply again the method described in the section 3.1 to provide single station estimations of the inversion parameters. When we have a large number of spectra contributing to the computation of source parameters, the addition of the site correction is not expected to significantly change the final estimates of the source parameters. However, since we corrected for systematic station residuals, the single station *MSE* is reduced and thus also the final uncertainty. While the inclusion of site effect in the described methodology can be delineated, its application to the dataset analyzed here does not allow to significantly reduce the MSE for single station due to the limited number of analyzed events per station. We would need to significantly increase the number of solutions per station to reduce the bias due to the site contributions, eventually decreasing the minimum magnitude of the dataset. However, this is beyond the scope of this study and deserves further investigation.

## 6. Conclusions

We developed a probabilistic framework based on the conjunction of states of information between data and model, to jointly retrieve earthquake source parameters and the anelastic attenuation factor. We modeled the observed far field displacement spectrum assuming a circular rupture model (Brune 1970) for the source and a Green's function characterized by a frequency-independent quality factor for the propagation. The forward operator is therefore defined on a set of 4 parameters: three parameters for the source – the seismic moment $M_0$, the corner frequency $f_c$ and the high-frequency decay exponent $\gamma$ – and one parameter – the Q-factor – for the propagation. These parameters are strongly correlated among each other. We estimated the joint probability density function (PDF) over the 4-D model space to extract the correlation matrix of the parameters; this allows to obtain estimates and uncertainties from the PDF, taking into account the correlations.

Since we modeled the observations with a non-linear operator, a global exploration of the model space is required in order to find the best solution to describe the data. We used a global optimization technique that relies on the definition of a Markov chain in the model space and on the combination of a deterministic minimization with a random exploration of the space (Wales & Doye 1997; Wales 2003).

In order to validate the developed methodology, we performed synthetic tests on spectra with different signal to noise ratios, defined on different frequency domains. The method proved its efficacy with all the synthetic cases analyzed here. The resolution of the estimates depends both on the SNR and the frequency bandwidth available for the inversion; at least 0.1 decades on the left and 0.4 decades on the right of the $f_c$ are required in the frequency domain to obtain reliable estimations for the inversion. Moreover, we showed that the uncertainty on the quality factor estimate depends on the value of $Q$ itself. Specifically, our tests show that when inverting for a constant $Q$ in an attenuation model in which the quality factor depends on the frequency, the retrieved $Q$ value might not represent an averaged value for the quality factor.

The method has been applied to the main events of the Central Italy 2016-2017 sequence ($M_L \geq 4.0$, 56 events). We found that the standard scaling $M_0 \propto \dfrac{1}{f_c^3}$ holds with an average static stress drop $\Delta\sigma = 2.1 \pm 0.3\ MPa$. For the main event ($M_W = 6.4 \pm 0.1$), we estimated a corner frequency $f_c = 0.15 \pm 0.03\ Hz$ and a high-frequency decay slope $\gamma = 2.14 \pm 0.08$; the source radius is $r = 8.3 \pm 1.8\ km$ and the static stress drop is $\Delta\sigma = 4 \pm 3\ MPa$.

The average stress drop retrieved in this study is consistent with the values inferred from the two major seismic sequences of the Central Italy region in the last two decades.

The Savage-Wood seismic efficiency ranges between 0.04 and 0.25, and it is almost constant with the explored seismic moment ($10^{14.16} - 10^{18.8}\ Nm$). Such low values may indicate an overshoot in the static final stress when compared to the dynamic frictional level.

**Figures**

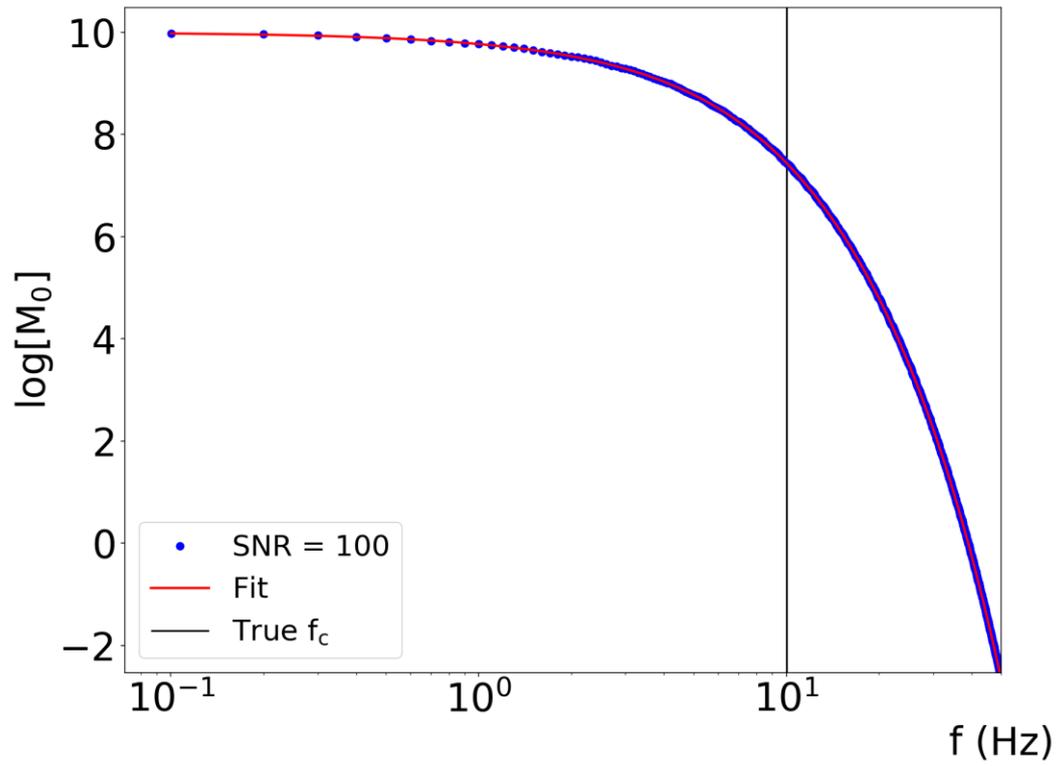

**Figure 1 :** Synthetic spectrum (blue curve) and solution retrieved from the inverse problem (red curve) for the case of SNR = 100.

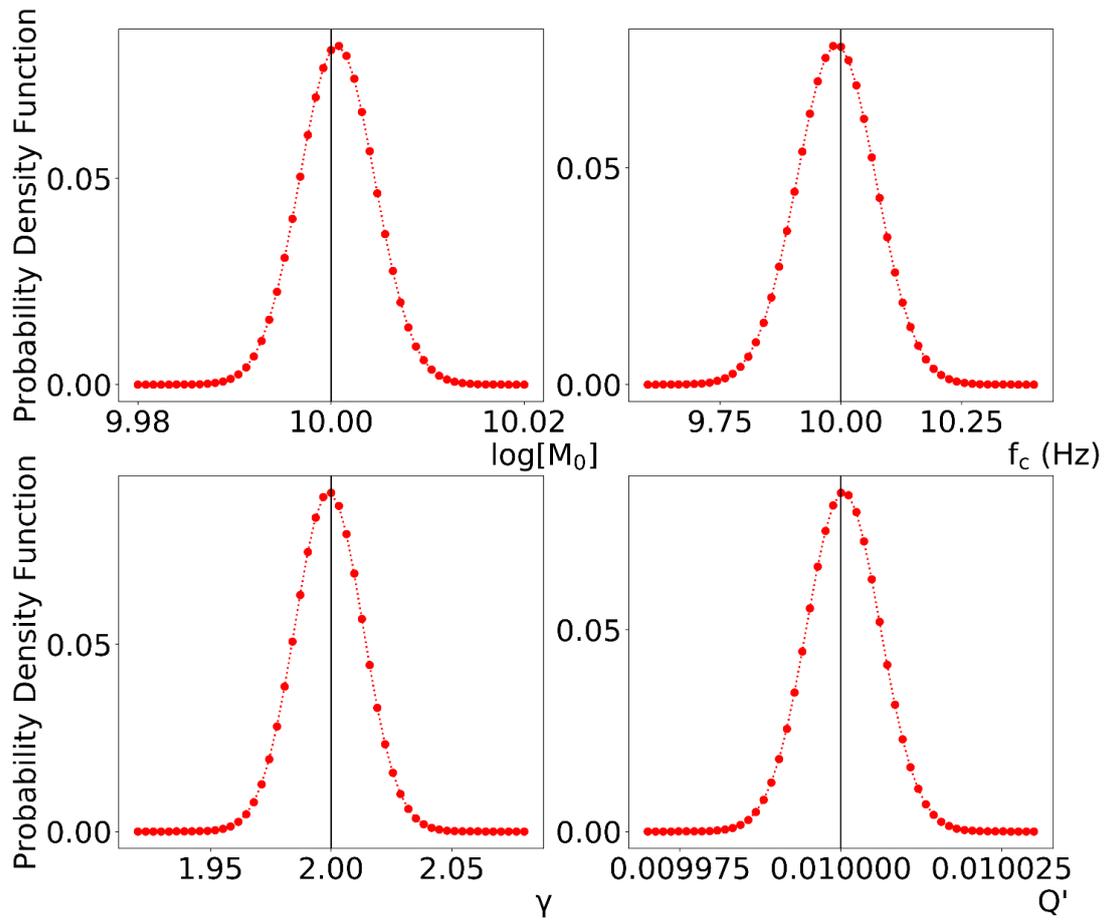

**Figure 2 :** 1-D marginal PDFs computed for a SNR = 100. In the four panels the PDF are represented for $\log(M_0)$, $f_c$, $\gamma$ and $Q'$. The black vertical line is the true value of the parameter. All the distributions show a Gaussian-like behavior.

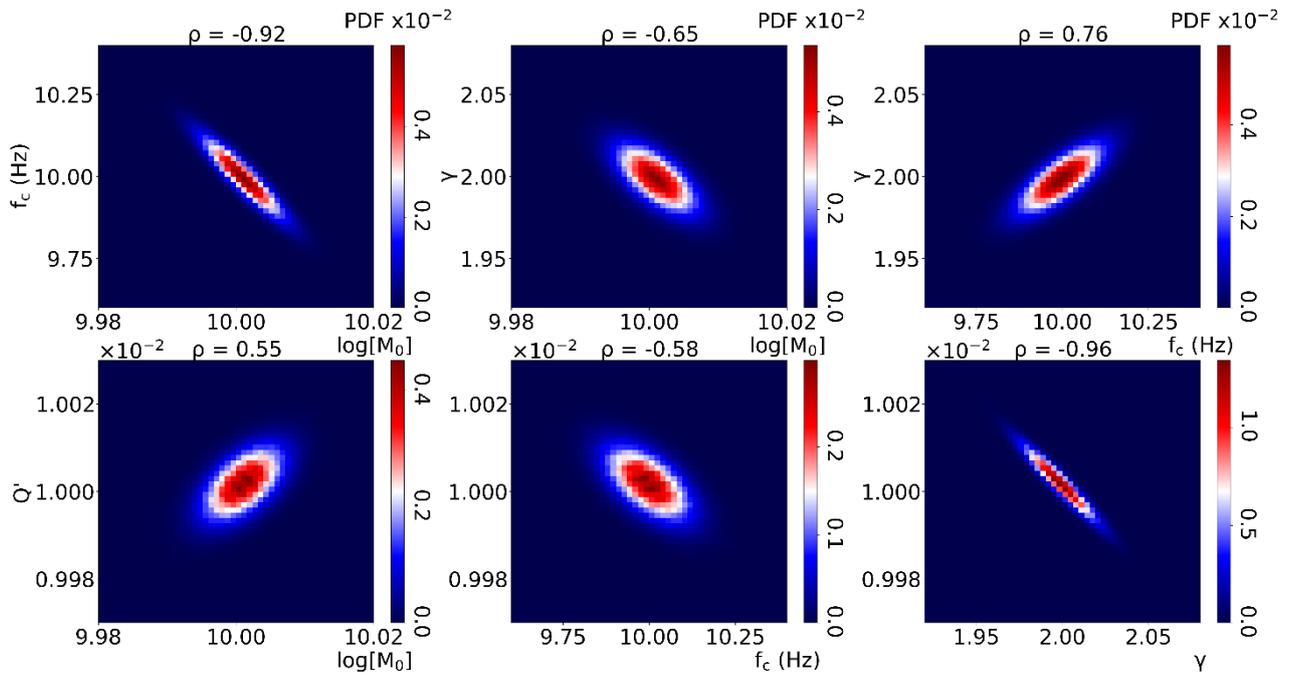

**Figure 3 :** 2-D marginal PDFs (heatmaps) computed for a SNR = 100. Correlation coefficients are at the top of each heatmap. We see that large anticorrelations occur for the couples log($M_0$)-$f_c$, and γ-Q', with correlation coefficients close to -1.

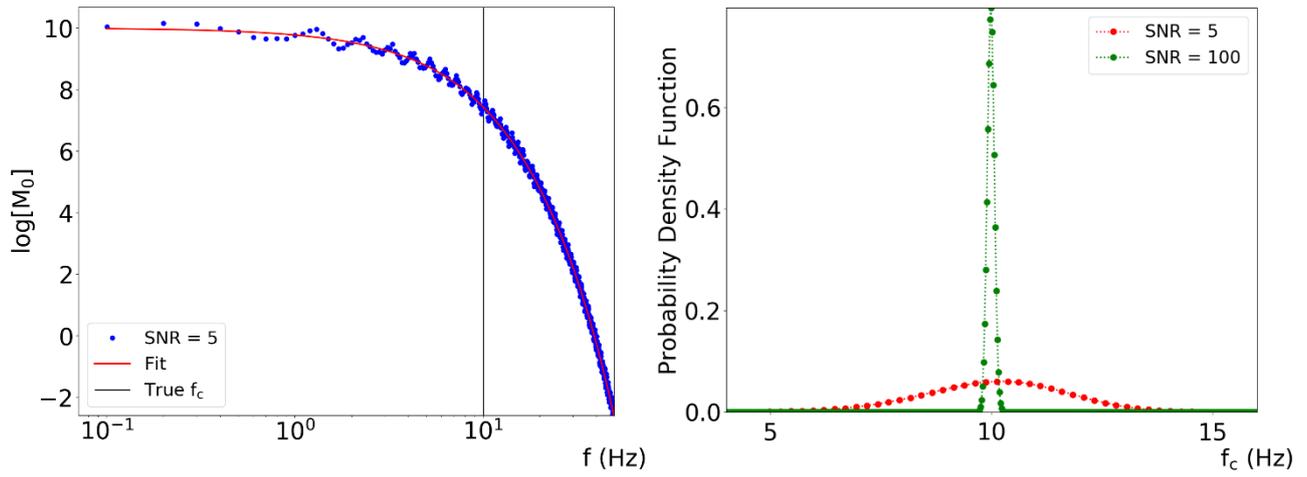

**Figure 4 :** Left panel : Synthetic spectrum generated with a SNR = 5 (blue curve) and the solution of the inverse problem (red curve). Right panel : comparison between two marginal PDFs as a function of $f_c$, for SNR = 100 (green curve) and SNR = 5 (red curve).

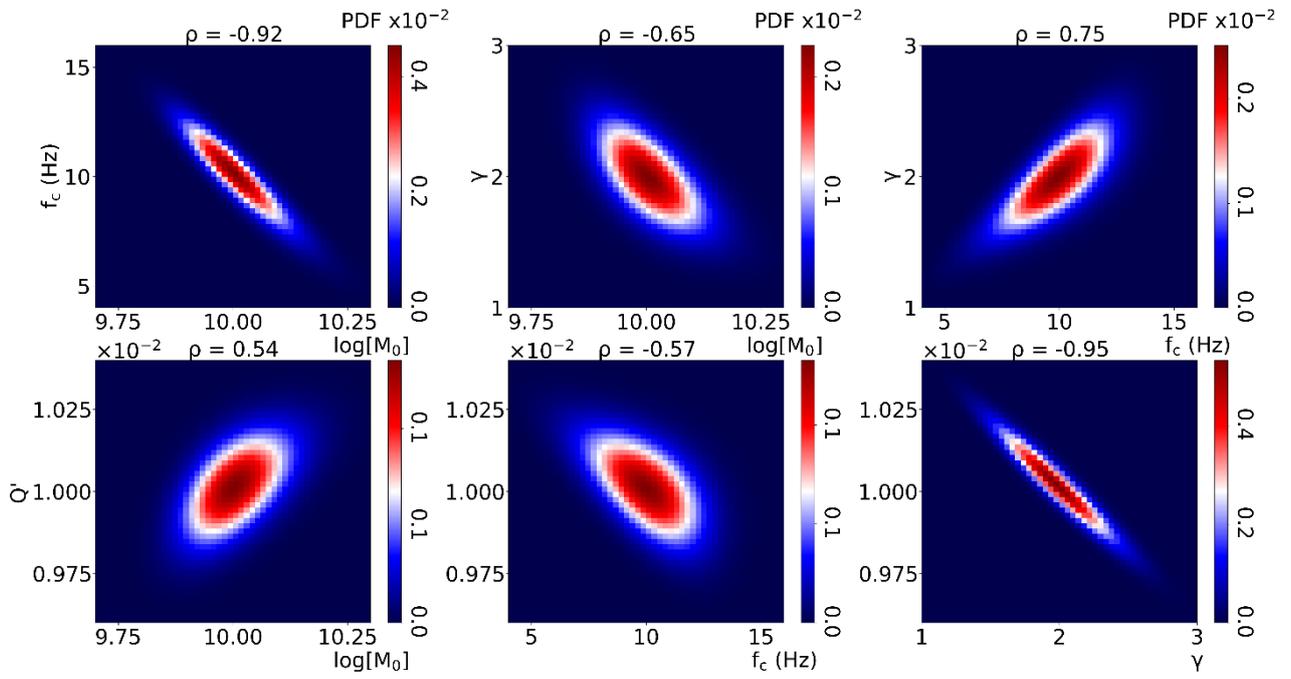

**Figure 5 :** 2-D marginal PDFs (heatmaps) computed for a SNR = 5. Correlation coefficients are at the top of each heatmap; both the shape and correlation coefficients do not significantly change as compared to the case of a SNR = 100 (Figure 3).

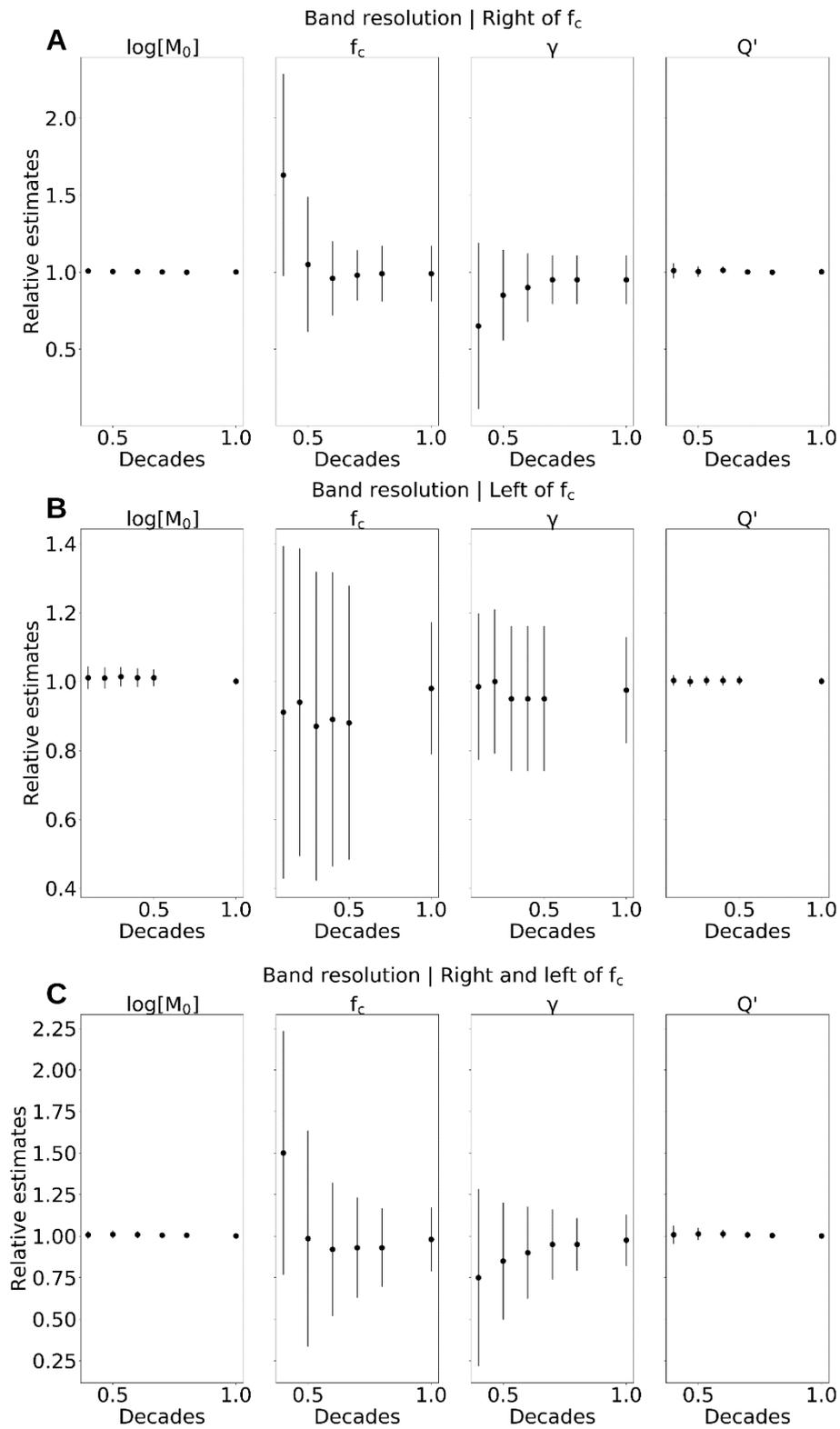

**Figure 6 :** Difference between the expected and the retrieved values for the parameters $\log(M_0)$, $f_c$, $\gamma$ and Q' as a function of the frequency bandwidth used for the inversion. Panels A, B and C corresponds to a change in the bandwidth from 0.4 to 1 decade to the right of $f_c$, to the left of $f_c$ and symmetrically around $f_c$, respectively.

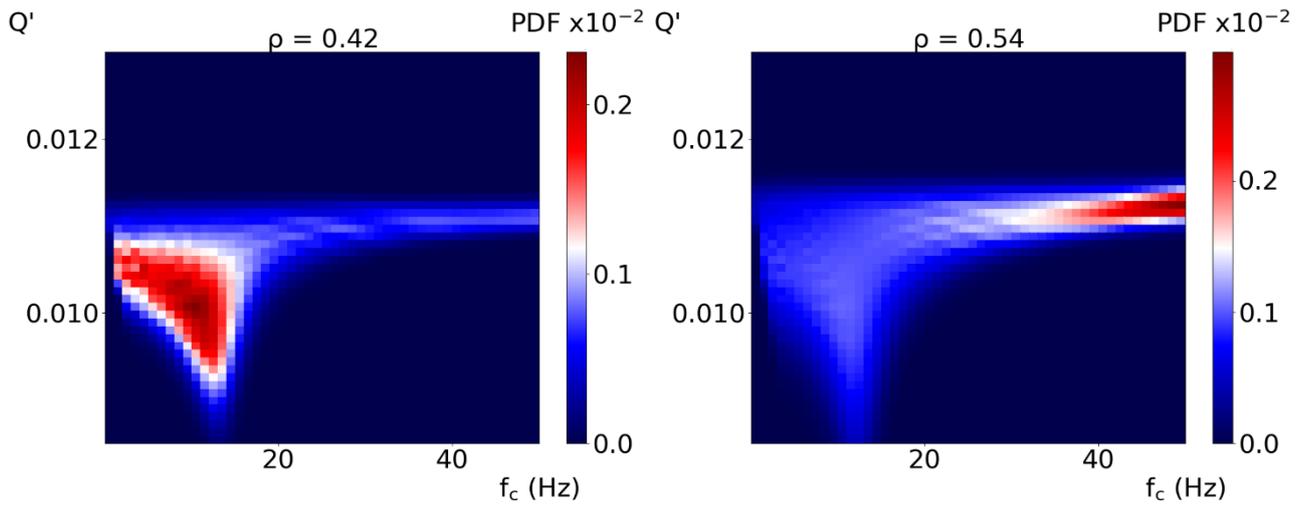

**Figure 7 :** 2-D marginal PDFs (heatmaps) for the parameters $f_c$ and Q'. In the left panel we represent the solution obtained using a bandwidth of 0.4 decade on the two sides of the corner frequency; in the right panel we show the solution with a bandwidth of 0.3 decade. We see that the maximum of the PDF migrates in the right panel to the upper limit of the explored $f_c$ range, far from the true value Q' = 0.1, $f_c$ = 10 *Hz*.

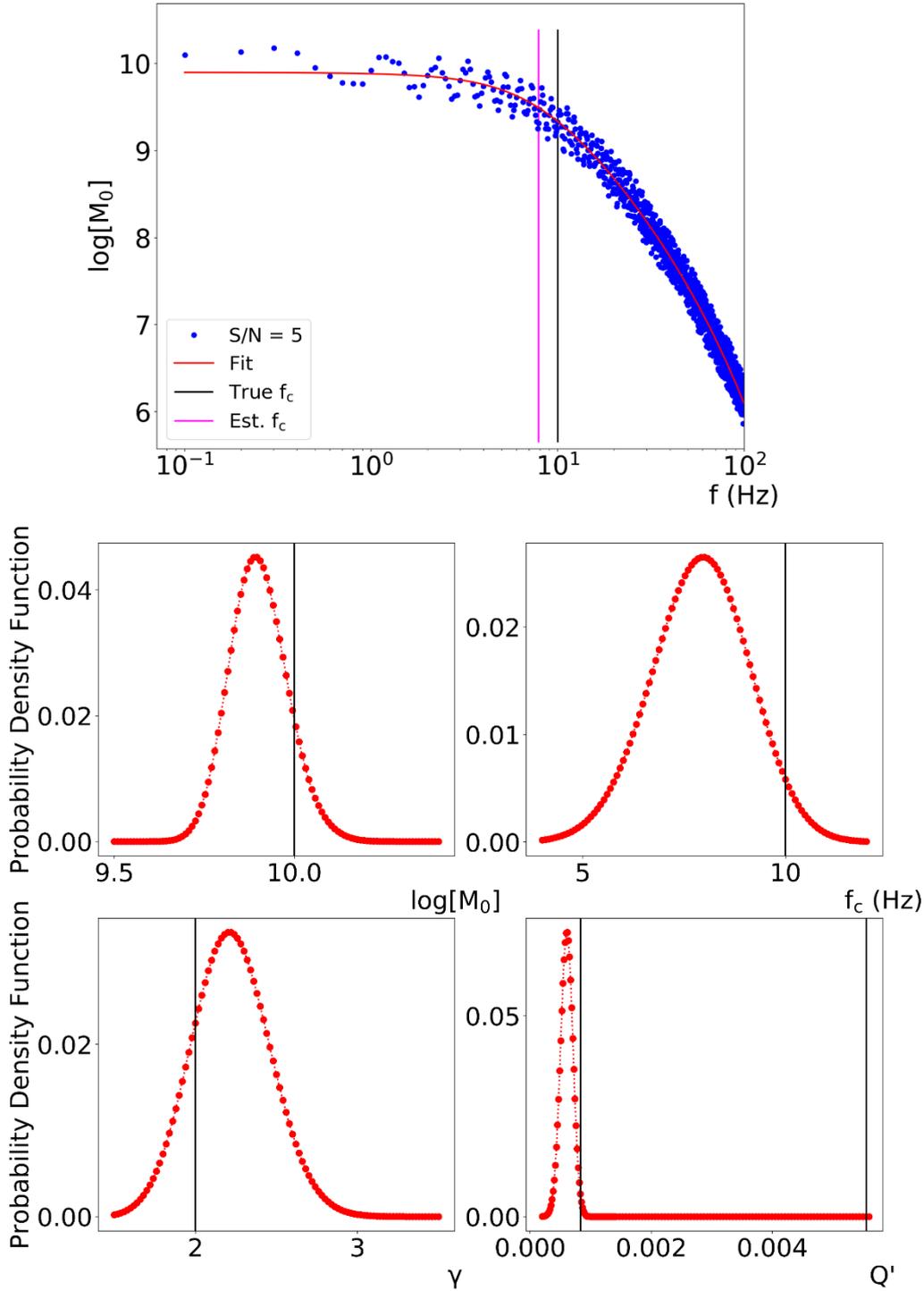

**Figure 8 :** Upper panel: Synthetic spectrum (blue curve) computed in an attenuation model with a frequency dependent $Q$ and solution retrieved from the inverse problem (red curve) using a constant quality factor. Lower panels: 1-D marginal PDFs for $\log(M_0)$, $f_c$, $\gamma$ and $Q'$. Black vertical lines point to the true value of the parameters or represent the bounds of $Q'$. All the distributions show a Gaussian-like behavior. Although the source parameters are well retrieved from the inversion, the mean value for $Q'$ is not included in the range of variability of the inverse of the quality factor.

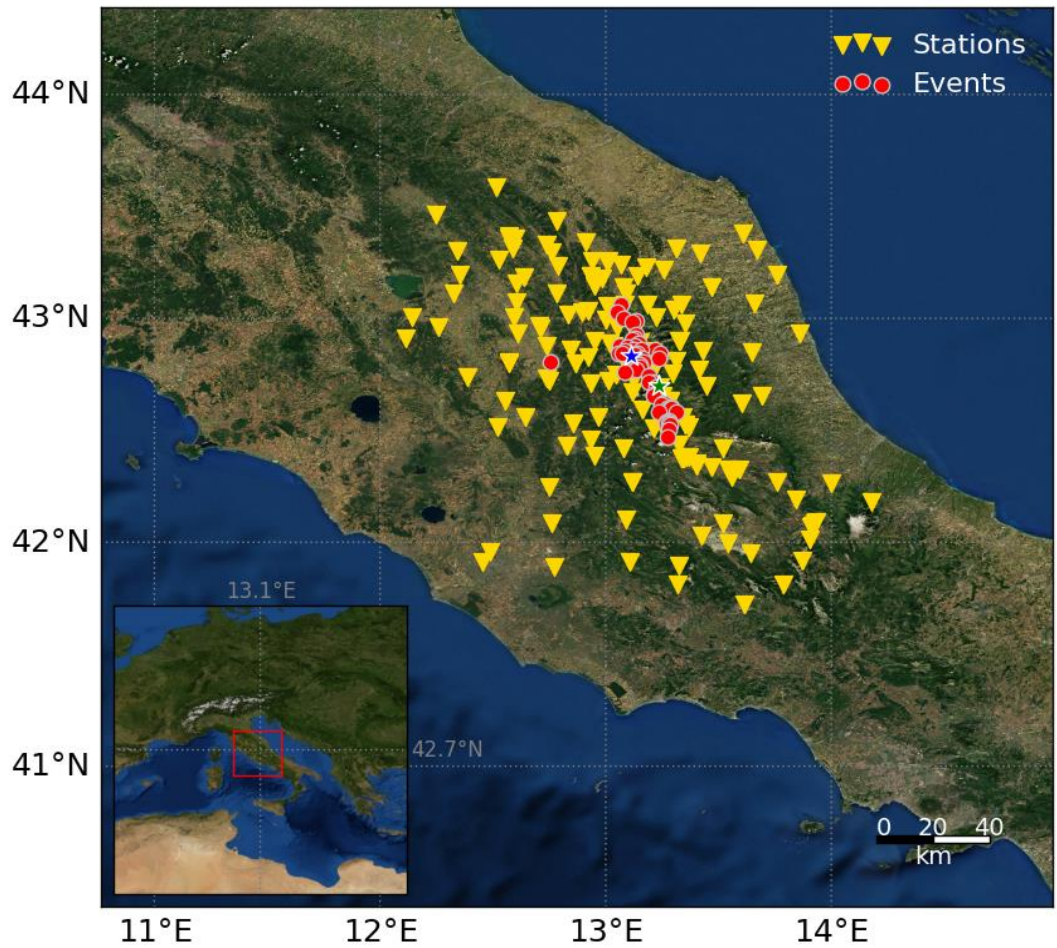

**Figure 9 :** Map representing the location of the events (red dots) and stations (yellow triangles). The green star and the blue star represent the locations of the $M_w$ 6.11 Amatrice earthquake and of the $M_w$ 6.4 Norcia earthquake, respectively.

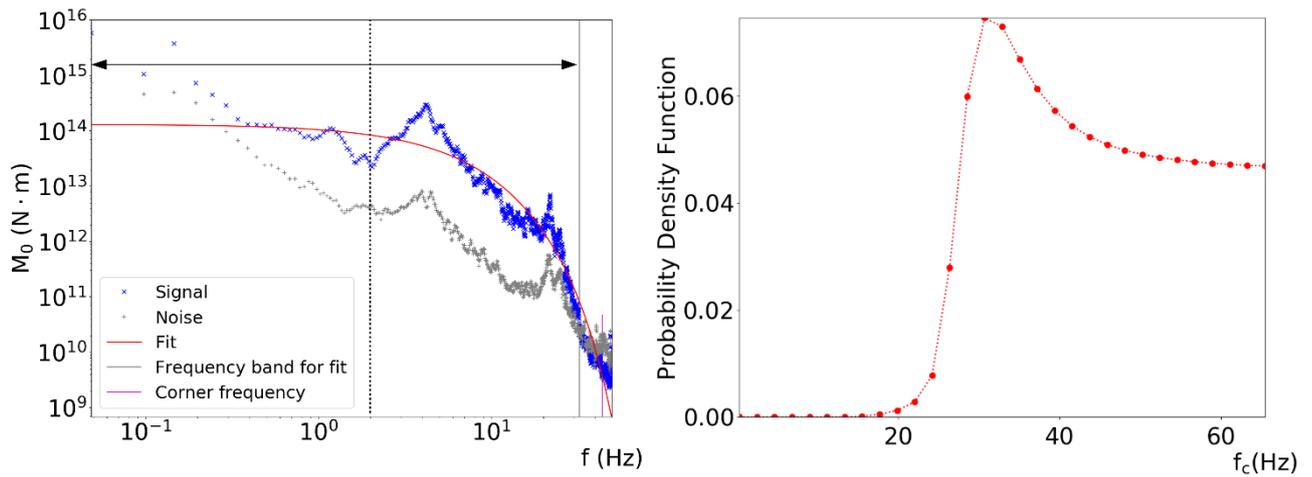

**Figure 10 :** Here we show an example of spectrum (Event ID-INGV: 7077781, Station: SSFR) that not provided a solution in terms of source parameters. In the left panel, we represent the spectrum of the signal (blue curve), the noise spectrum (gray curve) and the best-fit solution (red curve). In the right panel, we plot the 1-D marginal PDF for the parameter $f_c$. The solution is rejected because this marginal PDF does not overcome the threshold criterion for Gaussian similarity.

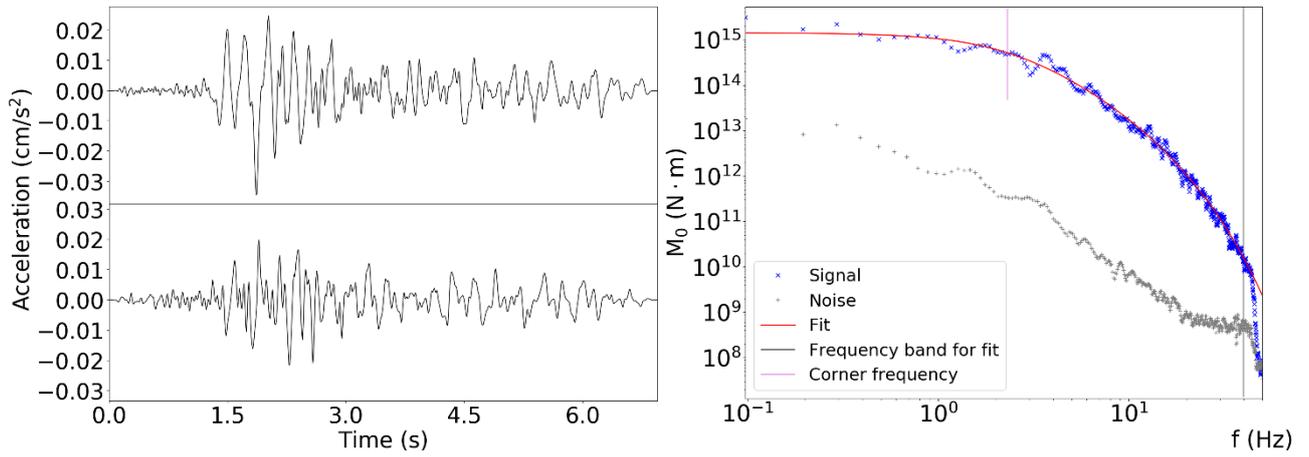

**Figure 11 :** An example of accepted solution. In the left panel we plot the 2 horizontal components of the signal selected for the analysis; in the right panel we represent the signal spectrum (blue curve), the noise spectrum (gray curve) and the best-fit solution (red curve). The traces are shown for the event ID-INGV 7141891 and the station FIAM. The time in the left panel is relative to 2016/08/24, 23:22:18.79 (UTC).

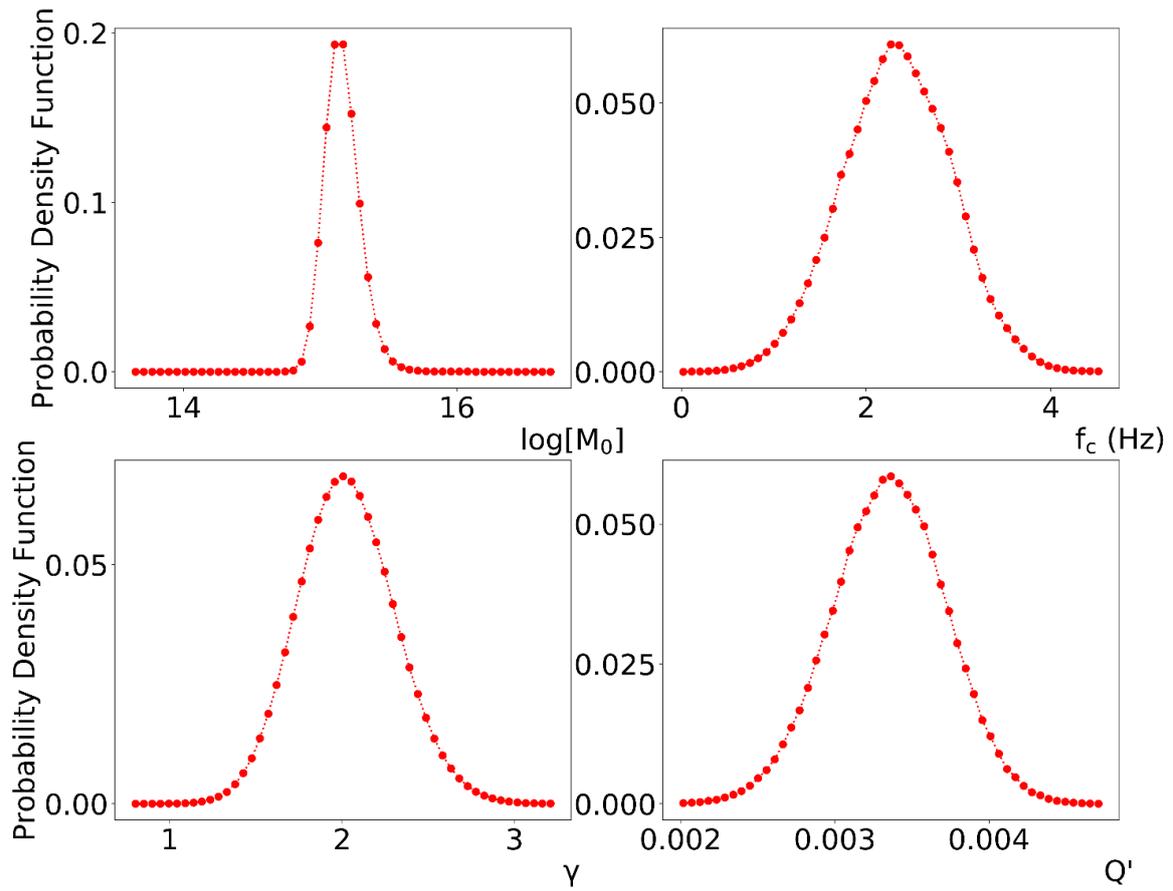

**Figure 12 :** 1-D marginal PDFs for the parameters $\log(M_0)$, $f_c$, $\gamma$ and $Q'$, for the same event-station as shown in Figure 11. All the curves show a Gaussian-like behavior.

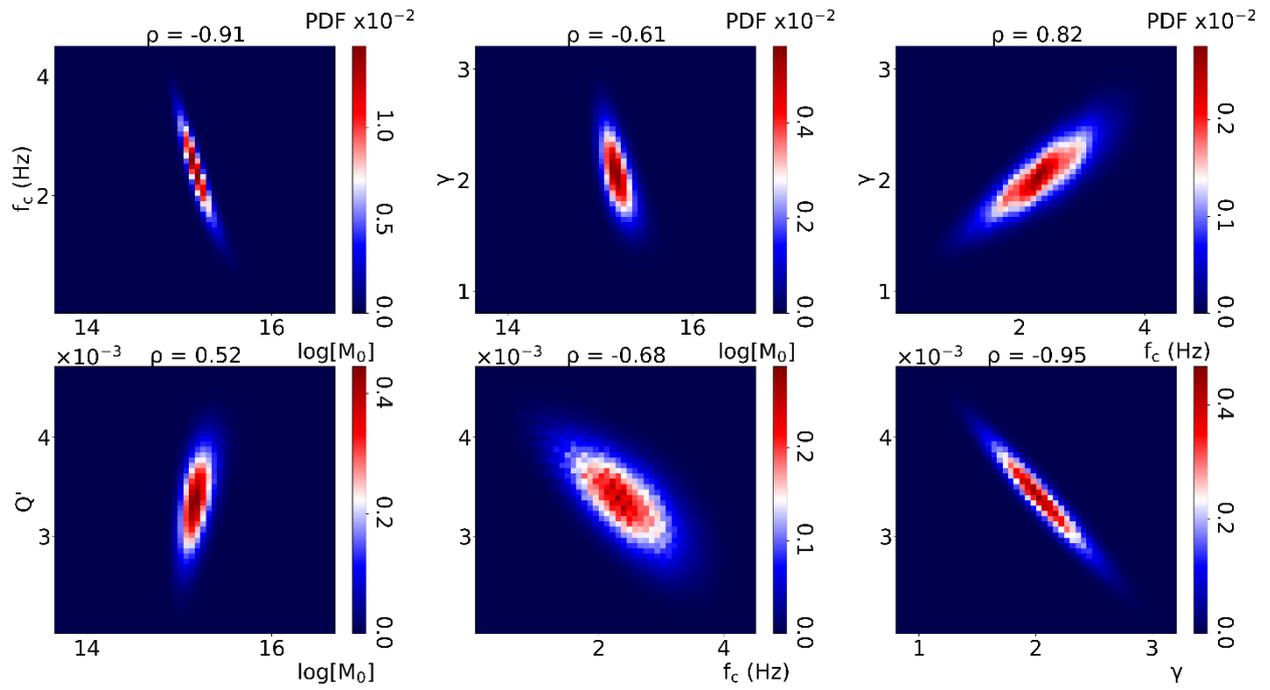

**Figure 13 :** 2-D marginal PDFs for the parameters log($M_0$), $f_c$, $\gamma$ and Q', for the same event-station as shown in Figure 11. These heatmaps look similar to the theoretical ones of Figure 3, indicating that correlations are mainly driven by model uncertainties.

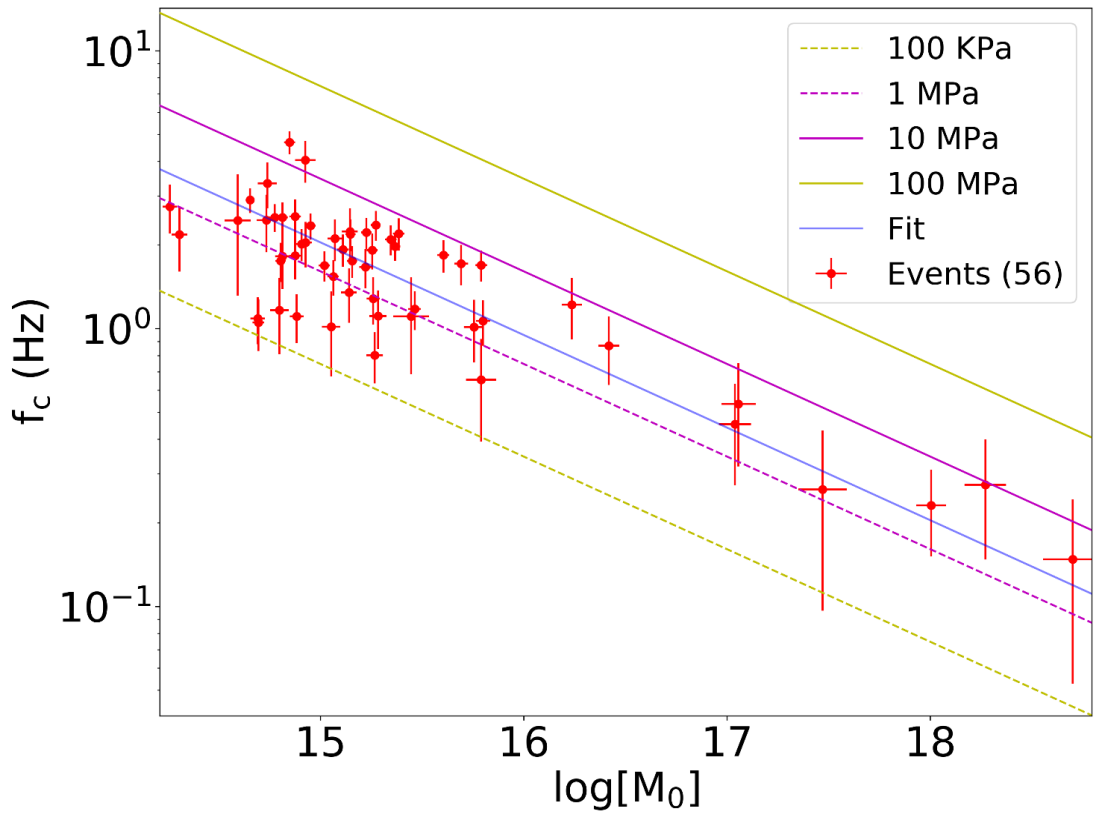

**Figure 14** : Scaling law between the corner frequency and the seismic moment. The red points are the solutions per event; the blue line is the best fit curve, with a slope of -3; the parallel line indicate the scaling with stress drops ranging from $\Delta\sigma = 0.1 MPa$ to $\Delta\sigma = 100 MPa$. The error bars are represented with a $3\sigma$ confidence level.

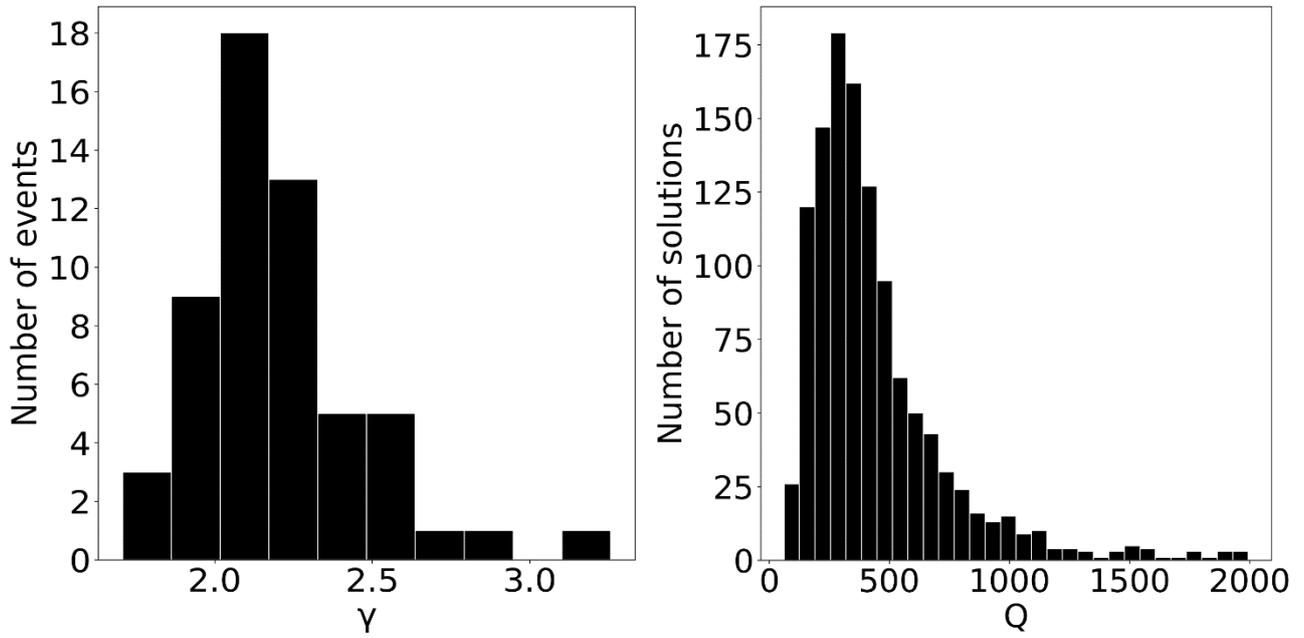

**Figure 15 :** Results for the source parameter γ and the quality factor Q. In the left panel we plot the histogram of the γ estimates per single event; in the right panel, we plot the histogram of the Q estimates as single station solution.

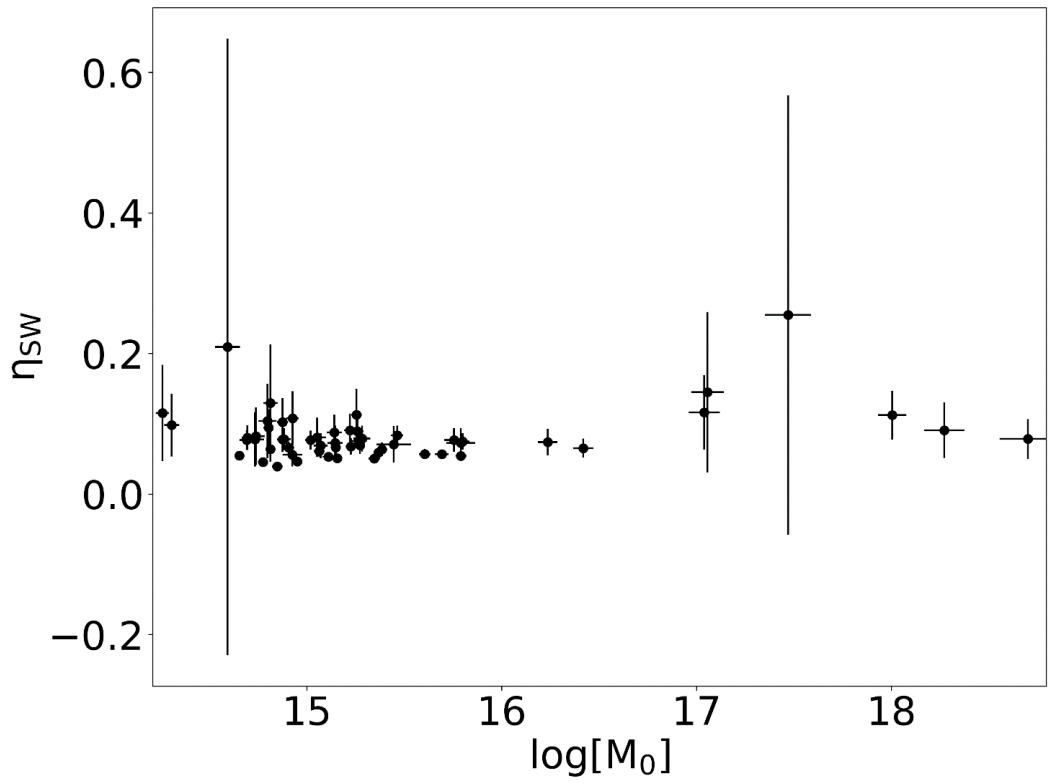

**Figure 16 :** Savage-Wood seismic efficiency versus seismic moment. The error bars are represented with a $3\sigma$ confidence level.

**Appendix A**

In the time domain, the displacement $u$ is the sum of the signal $s(t)$ and the noise $n(t)$ :

$$u(t) = s(t) + n(t) \tag{A1}$$

Here we prove that, for small $\dfrac{|\tilde{n}|}{|\tilde{s}|}$, the displacement amplitude spectrum can be written as :

$$\log(\tilde{u}) \approx \log(\tilde{s}) + \dfrac{\tilde{n}}{\tilde{s}} \cos(\varphi_S - \varphi_N) \tag{A2}$$

where $\tilde{s}$ and $\tilde{n}$ are the amplitude spectra and $\varphi_S$ and $\varphi_N$ the phase spectra of the signal and the noise, respectively; they are all functions of the frequency.

Applying the Fourier transform to the equation (A1), we have

$$\tilde{u}(f) = \tilde{s}(f) + \tilde{n}(f) \tag{A3}$$

It follows that :

$$|\tilde{u}| = \sqrt{(\tilde{s}+\tilde{n})(\overline{\tilde{s}}+\overline{\tilde{n}})} = \sqrt{|\tilde{s}|^2 + 2\operatorname{Re}(\tilde{s}\overline{\tilde{n}}) + |\tilde{n}|^2} = |\tilde{s}|\sqrt{1 + 2\dfrac{\operatorname{Re}(\tilde{s}\overline{\tilde{n}})}{|\tilde{s}|^2} + \dfrac{|\tilde{n}|^2}{|\tilde{s}|^2}}$$

$$\approx |\tilde{s}|\sqrt{1 + 2\dfrac{\operatorname{Re}(\tilde{s}\overline{\tilde{n}})}{|\tilde{s}|^2}} \approx |\tilde{s}|\left(1 + \dfrac{\operatorname{Re}(\tilde{s}\overline{\tilde{n}})}{|\tilde{s}|^2}\right) = |\tilde{s}| + \dfrac{\operatorname{Re}(\tilde{s}\overline{\tilde{n}})}{|\tilde{s}|} = |\tilde{s}| + |\tilde{n}|\cos(\varphi_S - \varphi_N) \tag{A4}$$

where we neglected the term $\dfrac{|\tilde{n}|^2}{|\tilde{s}|^2}$, and we approximate $\sqrt{1+x} \approx 1 + \dfrac{x}{2}$ for small $x$. The logarithm of (A4) gives the equation (A2) :

$$\log|\tilde{u}| = \log|\tilde{s}| + \log\left(1 + \dfrac{|\tilde{n}|}{|\tilde{s}|}\cos(\varphi_S - \varphi_N)\right) \approx \log|\tilde{s}| + \dfrac{|\tilde{n}|}{|\tilde{s}|}\cos(\varphi_S - \varphi_N) \tag{A5}$$

where $\log(1+x) \approx x$ for small $x$ .